\documentclass[onecolumn,12pt,draftcls]{IEEEtran}
\IEEEoverridecommandlockouts
\usepackage{float}
\usepackage{amsmath}
\usepackage{amsthm}
\usepackage{amssymb}	
\usepackage{graphicx}
\usepackage{subfigure}
\usepackage{enumitem}
\usepackage{algorithm}
\usepackage{stfloats}
\usepackage{cite}
\usepackage{cuted}
\usepackage{color}
\newcommand*{\J}{\jmath}
\newcommand{\diff}{\mathop{}\!d}
\usepackage{amsmath,amssymb,mathtools,bm,etoolbox}
\usepackage[center]{caption}

\DeclarePairedDelimiterXPP\Aver[1]{\mathbb{E}}{[}{]}{}{
	
	#1
}

\newtheorem{my_theorem}{Theorem}

\newtheorem{my_corollary}{Corollary}
\newtheorem{my_proposition}{Proposition}

\title{On the Performance of Multihop THz Wireless  System Over Mixed Channel Fading with Shadowing and  Antenna Misalignment}

\author{Pranay Bhardwaj,~\IEEEmembership{Graduate Student Member,~IEEE} and S.~M.~Zafaruddin,~\IEEEmembership{Senior Member,~IEEE}

	\thanks{The authors are with the Department of Electrical and Electronics Engineering, Birla Institute of Technology and Science, Pilani, Pilani-333031, Rajasthan, India; Email: \{p20200026, syed.zafaruddin\}@pilani.bits-pilani.ac.in.}
	
	 \thanks{ This work was supported in part by the Science and Engineering Research Board (SERB), Government of India, under MATRICS Grant MTR/2021/000890 and Start-up Research Grant SRG/2019/002345.}	
}
\begin{document}
\maketitle
	\begin{abstract}
The existing relay-assisted terahertz (THz) wireless system is limited to dual-hop transmission with pointing errors and short-term fading without considering the shadowing effect. This paper analyzes the performance of a multihop-assisted backhaul communication mixed with an access link under the shadowed fading with antenna misalignment errors. We derive novel probability density and distribution functions of the signal-to-noise ratio (SNR) of the multihop link employing channel-assisted (CA)  and fixed-gain (FG) amplify-and-forward (AF) relaying for each hop by considering independent but not identically distributed (i.ni.d) $\alpha$-$\mu$ fading channel with pointing errors.  We present analytical expressions for the outage probability and average bit-error-rate (BER) and develop asymptotic analysis  in high SNR regime to derive the diversity order for both CA-multihop and FG-multihop backhaul links. We use the derived results of the multihop systems to analyze the statistical performance of the integrated backhaul-access relaying for both uplink and downlink transmission considering the generalized-$K$ shadowed fading model in the access link. We also present exact and asymptotic expressions of the outage probability for the considered mixed system. We perform computer simulations to provide design and deployment aspects of employing multiple relays to enhance THz wireless transmissions.

\end{abstract}
			
\section{Introduction}
Terahertz (THz) communication is emerging as a viable technology for 6G wireless networks. The THz band ($0.3$ to $3$ THz) has a large unlicensed bandwidth to deliver massive data rates with low latency and improved security \cite{Dang_2020_nature,Akyildiz_2020, Song_2011_THz,Rappaport_2019,Sarieddeen_2020_THz_survey,Koenig_2013_nature}.   High-speed THz connectivity can be set in hard-to-access terrains more efficiently  compared to the infrastructure-based fiber-optic wireline link.  Although THz wireless transmission has a great potential to be a major disruptive technology,  it suffers from higher atmospheric attenuation caused by molecular absorption, hardware impairments,  and misalignment errors when the transmit and receive antennas do not align for line-of-sight (LOS) transmissions limiting the physical range of communication significantly. Hence, it is desirable to devise methods for extending the communication range of THz wireless systems under adverse propagation conditions such as higher path loss, misalignment errors, and the shadowing effect between the source and destination.

Cooperative relaying is a potential technique to extend the communication range and improve the quality of system performance.  The popular amplify-and-forward (AF) and decode-and-forward (DF) relaying schemes have been extensively investigated for  radio-frequency (RF) \cite{Hasna_2003_multihop_nakagami_letter,Karagiannidis_2006_multihop_nakagami,Karagiannidis_2006_multihop_nakagami_2,Morgado_2010_multihop_nakagami,Lateef_2011_multihop_nakagami,Trigui_2011_multihop_nakagami,Cao_2010_multihop_DF_generalisedk,Lin_2015, Badarneh_2016_multihop_DF, Dixit_2017_multihop_kmu_DF} and free-space optics (FSO) \cite{Tsiftsis_2006_multihop_AF_FSO,Tang_2014_multihop_AF_FSO_Pointing,Zedini_2015_multihop_AF_FSO_pointing,Ashrafzadeh_2020_multihop_FSO,Wang_2016_multihop_generalised_PE,Issaid_2017_multihop_generalised_PE,Alheadary_2017} over various fading scenarios.  Fixed-gain (FG) and channel-assisted (CA) are two practical approaches for AF relaying. In the FG relaying, the statistical information of the received signal from the previous hop is used to control the gain of the relay. Moreover, the near-optimal CA relaying computes the gain by inverting the channel without considering the effect of noise. 

Recently, there has been an increased research interest for mixed dual-hop THz relaying systems \cite{Boulogeorgos_Error, Pranay_2021_VTC,Pranay_2021_TVT,Pranay_2021_letters, Boulogeorgos_2020_THz_THz,ants2021_pranay, Li_2021_THz_AF}. In \cite{Boulogeorgos_Error}, the average bit-error-rate (BER) performance of a DF-assisted THz-RF link in a backhaul-broadband access framework is studied over $\alpha$-$\mu$ fading.  The $\alpha$-$\mu$  is a generalized short-term fading model which includes Weibull, negative exponential, Nakagami-$m$, and Rayleigh distribution as a special case \cite{Yacoub_alpha_mu}. Recent  experiment campaigns validates  $\alpha$-$\mu$ distribution for the short term fading in THz link at a $152$ \mbox{GHz} carrier frequency for a link length within $50$\mbox{m} \cite{Papasotiriou2021,scientific_report21}. In our recent paper \cite{Pranay_2021_TVT}, we developed analytical expressions of average signal-to-noise ratio (SNR), ergodic capacity, outage probability, and average BER considering asymmetric fading conditions for both THz and RF links with real-valued $\mu$. To extend the range of THz transmissions, dual-hop relaying has been investigated \cite{Boulogeorgos_2020_THz_THz, ants2021_pranay, Li_2021_THz_AF}.  In \cite{Boulogeorgos_2020_THz_THz,ants2021_pranay}, the authors analyzed the performance of a dual-hop THz-THz backhaul system employing the DF protocol with  $\alpha$-$\mu$ fading model with pointing errors for both the links. Considering similar fading distributions,  the FG  relaying was studied for the dual-hop THz-THz system \cite{Li_2021_THz_AF}. 

In the aforementioned and related research, relaying is limited to dual-hop, and the access link operates over conventional RF frequencies. It is expected that the high bandwidth THz transmissions can be employed for 6G radio access network (RAN). Although the THz radio is immune to atmospheric turbulence and adverse weather conditions than the FSO technology, it is susceptible to human movement between the transmitter and the receiver leading to dynamic shadowing \cite{Giorgos2020, Ye_2021_generalized_k}. In \cite{Ye_2021_generalized_k}, the generalized-$K$ fading distribution is adopted to model the  THz transmissions in the access link. However, the effect of pointing errors is not considered in \cite{Ye_2021_generalized_k} for a better estimate of the THz performance. Moreover, the multihop THz system extending the range of backhaul link has not been studied in the related literature to the best of the authors' knowledge. It is well-known that analyzing the performance of multihop transmissions, even for single-parameter RF fading distributions is challenging, especially when multihop links are assisted by the AF relaying. In general,  the performance of multihop-assisted AF relaying is approximated by expressing the resultant SNR  as the product of SNR of individual links using the inequality of harmonic and geometric mean. Further, the method of induction (generalization for $N$ hops realized through expressions derived for the first few hops) is applied to derive the probability density function (PDF), and thereby cumulative distribution function (CDF) of the multihop system
\cite{Tang_2014_multihop_AF_FSO_Pointing}. Note that the mathematical induction method is not applicable for THz transmissions under the combined effect of $\alpha$-$\mu$ fading with pointing errors and requires novel approaches for its statistical analysis. Moreover, it is desirable to derive exact analytical expressions of a  CA relaying,  which serves as a benchmark for all practical multihop systems employing AF relays \cite{Tang_2014_multihop_AF_FSO_Pointing}. 

In this paper, we analyze the performance of a multihop-assisted backhaul mixed with a shadowed access link over the THz band for both uplink and downlink transmissions. We consider the AF relaying for the multihop system and employ DF and AF protocols to interface with the radio access link for uplink and downlink, respectively.   We consider the generalized i.ni.d. $\alpha$-$\mu$ fading distribution for the multihop link and the composite generalized-$K$ shadowed fading for the radio access link, both under the combined effect of pointing errors.  
We list the major contributions of the paper as follows:
\begin{itemize}[leftmargin=*]
	\item We derive novel  PDF and CDF  of the end-to-end SNR  of the multihop backhaul link  employing  CA relaying for each hop.
	\item  We develop an  upper bound for the FG relaying based multihop transmission by deriving the exact PDF of the product of $N$ i.ni.d. random variables that are distributed according to the product of  $\alpha$-$\mu$ fading and pointing errors.

 \item Using the derived statistical results, we analyze the outage probability and average BER performance for CA-multihop and FG-multihop systems. We also analyze the multihop transmissions for some special cases of the THz channel and present an asymptotic analysis in high SNR regime for a better insight into the system behavior. 

\item We develop statistical results of an all FG-relayed downlink transmission by employing the FG protocol in the last hop to mix the FG-multihop and access link in terms of the bivariate Fox's H-function. We analyze the $N+1$-hop downlink system by deriving upper bounds on the outage probability and average BER. Note that our derived analytical expressions become exact for a single-relay ($N=1$) system resulting in a typical dual-hop mixed transmission. We also analyze the performance of uplink transmission by employing the DF relaying protocol in the first hop to mix the access link and CA-multihop.
\item We  derive asymptotic expressions of the outage probability and average BER in high SNR region and develop diversity order to provide design aspects of channel and system parameters for the integrated backhaul-access relaying  transmissions.  
\item We validate the derived analysis using Monte-Carlo simulations and demonstrate the significance of multihop relaying to extend the communication range for THz wireless transmissions under the effect of fading, pointing error, and shadowing.
\end{itemize}

\subsection*{Related Works}
There has been increased research interest in evaluating the performance of relay-assisted THz wireless systems considering realistic propagation environments \cite{Abbasi_2017,Rong_2017,Xia_2017,Xia_2020_ICC,Boulogeorgos_2020_THz_THz,Boulogeorgos_Error,Pranay_2021_TVT,Pranay_2021_letters,Mir_2020,Li_2021_THz_AF}. The dual-hop THz transmission was studied in \cite{Boulogeorgos_Error, Pranay_2021_VTC,Pranay_2021_TVT,Pranay_2021_letters, Boulogeorgos_2020_THz_THz, Li_2021_THz_AF}. In \cite{Mir_2020}, the authors presented a hybrid precoding design for the dual-hop multiple-input multiple-output (MIMO) THz network. However, there is limited research in multihop THz communications. In \cite{Xu_2016}, an energy-efficient multihop routing protocol for the THz wireless nano-sensor network has been proposed. The authors in \cite{huang_2021_multihop_RIS_THz} investigate the beam-forming techniques for multihop reconfigurable intelligent surfaces (RIS) empowered THz communications over Rician fading.

\begin{table}[tp] 
	
	\caption{Literature on relay-assisted THz wireless systems}	
	\label{table2}
	\begin{center}
		\begin{tabular}{|p{1.4cm}|p {3.5cm}|p {4.6cm}|p {6cm}|}
			\hline   
			\textbf{Reference}  & \textbf{Relaying Model} & \textbf{Relaying Protocol} &  \textbf{Fading Model}  \\ 
			\hline  
			\cite{Boulogeorgos_Error} & Mixed THz-RF &Dual-hop DF  &  i.i.d. $\alpha$-$\mu$ model for both the links. \\  
			\hline
			\cite{Pranay_2021_TVT} & Mixed THz-RF &  Dual-hop DF  & i.ni.d. $\alpha$-$\mu$ model for both the links.  \\
			\hline
			\cite{Pranay_2021_letters}  & Mixed THz-RF &  Dual-hop FG  & $\alpha$-$\mu$ model for THz link and $\alpha$-$\kappa$-$\mu$ shadowed fading for the RF. \\
			\hline
			\cite{Li_2021_THz_AF} &   THz-THz &  Dual-hop FG & i.ni.d. $\alpha$-$\mu$ model for both the links.  \\
			\hline
			This paper &  Multihop THz & CA-multihop and FG-multihop & i.ni.d. $\alpha$-$\mu$ model for each hop. \\
			 & Mixed Multihop THz-THz & Uplink with DF;  Downlink with FG& i.ni.d. $\alpha$-$\mu$  and generalized-$K$ fading. \\
			\hline
		\end{tabular}
	\end{center}
\end{table}

On the other hand, there has been an extensive research on the AF-assisted multihop transmission for RF systems \cite{Hasna_2003_multihop_nakagami_letter,Karagiannidis_2006_multihop_nakagami,Karagiannidis_2006_multihop_nakagami_2,Morgado_2010_multihop_nakagami,Lateef_2011_multihop_nakagami,Trigui_2011_multihop_nakagami,Lin_2015} and FSO systems with pointing errors \cite{ Tsiftsis_2006_multihop_AF_FSO,Tang_2014_multihop_AF_FSO_Pointing,Zedini_2015_multihop_AF_FSO_pointing,Ashrafzadeh_2020_multihop_FSO,Issaid_2017_multihop_generalised_PE,Alheadary_2017}. The authors in \cite{Hasna_2003_multihop_nakagami_letter} and \cite{Tsiftsis_2006_multihop_AF_FSO} analyzed the exact outage probability of  CA-multihop transmissions over Nakagami-$m$ and Gamma-Gamma fading channels, respectively. In \cite{Karagiannidis_2006_multihop_nakagami}, the authors used the method of induction to derive bounds on the performance of CA and FG assisted multihop transmissions over Nakagami-m fading channels. Similarly, the method of induction was used to derive bounds on the performance of CA and FG-assisted multihop transmissions for FSO systems with Gamma-Gamma atmospheric turbulence and pointing errors \cite{Tang_2014_multihop_AF_FSO_Pointing}. In \cite{Ashrafzadeh_2020_multihop_FSO}, an exact PDF of FG-assisted multihop transmission for an FSO system with double generalized Gamma with pointing errors was presented by generalizing the analysis for $N$ hops using the results obtained up to three hops. To this end, it should be mentioned that there is an extensive literature on the DF-based multihop relaying \cite{Cao_2010_multihop_DF_generalisedk, Badarneh_2016_multihop_DF,Dixit_2017_multihop_kmu_DF,Wang_2016_multihop_generalised_PE,Issaid_2017_multihop_generalised_PE, Yi_2019}. However, the DF system requires each hop's channel state information (CSI) for decoding. In \cite{Yi_2019}, the DF relaying was used in each hop for the mixed system consisting of multihop-assisted FSO and a single RF link.

In the light of related literature, it is desirable to realize a mixed wireless link consisting of backhaul/fronthaul and broadband access network both over THz wireless technology where each hop of transmission being integrated using low complexity relaying protocols. It should be mentioned that the propagation characteristics and application scenarios for THz are different from other state-of-the-art technologies such as RF and FSO.

\emph{Notations}: The notation $(.)_N$ and $(.)_A$ denotes the $N$-hop backhaul link and the access link, respectively. $\Aver.$ indicates the average value of a random variable and $\Gamma(a,t)=\int_{t}^{\infty} s^{a-1}e^{-s}ds$ denotes the upper incomplete Gamma function. $G_{p,q}^{m,n}(.|.)$ and $H_{p,q}^{m,n}(.|.)$ denotes Meijer's G and Fox's H-functions, respectively.   

\section{System Model}\label{sec:system_model}
\begin{figure}[tp]	
	\centering
	\includegraphics[width=\columnwidth]{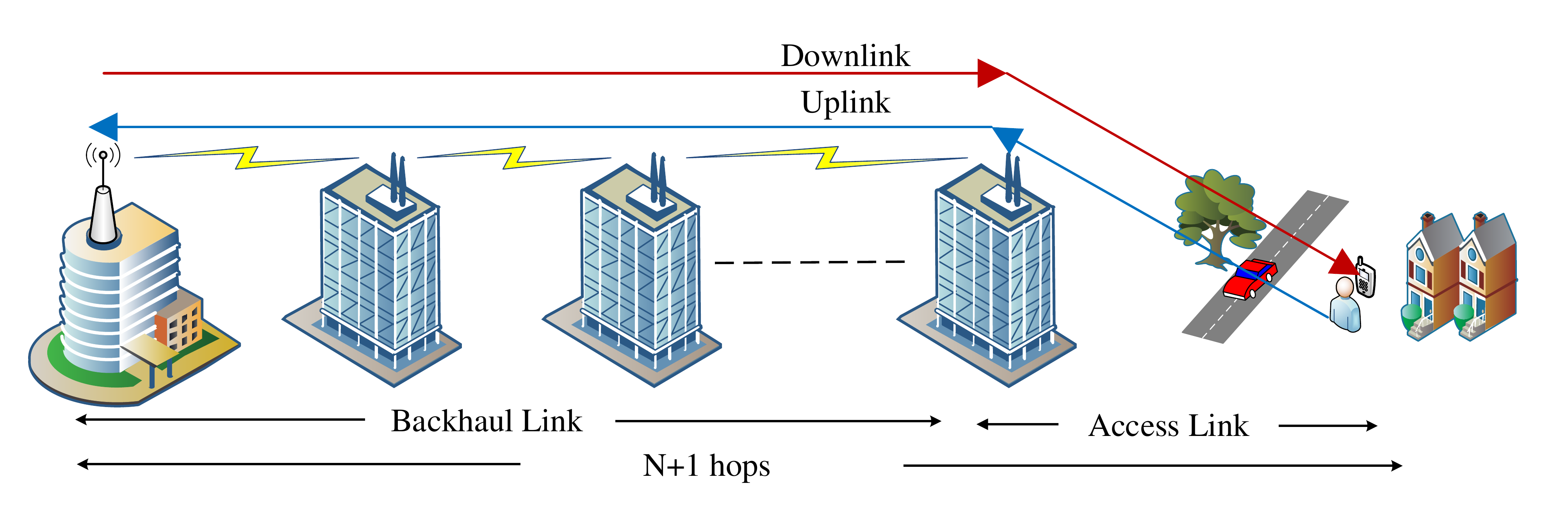}
	\vspace{-6mm}
	\caption{Multihop relay-assisted backhaul transmissions integrated with shadowed access link.}
	\label{fig:system_model}
\end{figure}
We consider a multihop-assisted backhaul transmission mixed with a broadband access link, both operating in the THz band. The transmission system consists of $N$+$1$ hops from the source ($S$) to the destination ($D$), where $N$ hops correspond to the backhaul connectivity and a single link of the access network, as shown in Fig.~\ref{fig:system_model}. We consider both uplink and downlink scenarios: signal transmission from the access node to the backhaul in the uplink and backhaul to the access node in the downlink transmission.

We consider i.ni.d. $\alpha$-$\mu$ distributed fading  parameters $\{\alpha_i, \mu_i\}_{i=1}^N$ combined with pointing errors and deterministic path loss to model the channel of each hop of the backhaul THz link. Using AF relaying protocol and denoting $G_i$ as the gain of $i$-th relay with $G_0=1$, the received signal at the $N$-th relay is given by \cite{Tang_2014_multihop_AF_FSO_Pointing}\footnote{There is a typo error in equation (17) \cite{Tang_2014_multihop_AF_FSO_Pointing}. In the second term of R.H.S., the product range should be from $j=i$ to $j=N-1$.}:
\begin{eqnarray} \label{eq:multihop_rcvd_signal_eqn}
	y_N = \prod_{i=1}^{N} G_{i-1} h_i x + \sum_{i=1}^{N-1} w_i \left(\prod_{j=i}^{N-1}G_j h_{j+1}\right) + w_N 
\end{eqnarray}
where $x$ is the transmitted signal with power $P_i$ and $w_i$ is the additive white Gaussian noise (AWGN) of the $i$-th relay with variance $\sigma^2_{wi}$ and $w_N$ is the AWGN of the $N$-th relay. The channel coefficient of the $i$-th hop is given as $h_{i} = h_{l_i} h_{p_i}h_{f_i}$, which includes the path gain $h_{li}$, short-term fading $h_{fi}$, and pointing errors $h_{pi}$. The path gain for the $i$-th link is $h_{li} = \frac{c\sqrt{G_{t}G_{r}}}{4\pi f d_i} \exp\big(-\frac{1}{2}k_id_i\big)$, where $c$ is the speed of light, $G_{t}$ and $G_{r}$ are the gains of transmitting and receiving antennas, respectively, $f$ is the frequency of operation, $k_i$ is the absorption coefficient and $d_i$ denotes the link distance of the $i$-th hop such that $d_1= \sum_{i=1}^{N} d_i$ is the total distance of the backhaul link.

We use the zero boresight statistical model for the pointing errors $h_{pi}$ with PDF \cite{Farid2007}:
\begin{eqnarray} \label{eq:pdf_hp}
	f_{h_{p_i}}(h_p)= \frac{\phi_i^2}{S_i^{\phi_i^2}}h_{p}^{\phi_i^{2}-1},0 \leq h_{p} \leq S,
\end{eqnarray}
where  $S_i$ and $\phi_i$ depict pointing error parameters for the $i$-th power. Here, the term $S_i$ denotes the fraction of collected power when the transmitter and receiver antennas are perfectly aligned, and $\phi_i$ denotes the ratio of normalized beam-width to the standard deviation of the jitter.

Using \eqref{eq:pdf_hp} and the PDF of $\alpha$-$\mu$ distribution \cite[eq. (1)]{Yacoub_alpha_mu}, the PDF of  $h_{p_if_i}=h_{pi}h_{fi}$, which combines the effect of short-term fading and pointing errors is presented in  \cite{Boulogeorgos_Error}:	
\begin{equation} \label{eq:pdf_hpf}
	f_{|h_{pf_i}|}(x) = A_i  x^{\phi_i-1} \Gamma (B_i, C_ix^{\alpha_i} )
\end{equation}	
where $A_i = \frac{\phi_i S_i^{-\phi_i} \mu_i^{\frac{\phi_i}{\alpha_i}}}{\Omega_i^{\phi_i} \Gamma(\mu_i)} $, $ B_i = \frac{\alpha_i \mu_i - \phi_i}{\alpha_i} $ and $C_i =  \frac{\mu_i}{\Omega_i^{\alpha_i}} S_i^{-\alpha_i}$. Here, $\alpha_i$ and $\mu_i$  are the distribution parameter of the $i$-th link. $\Omega_i$ is the $\alpha_i$ root-mean value of the fading channel envelope. We denote instantaneous SNR of the $i$-th link as $\gamma_{i} = \gamma_i^0 |h_{i}|^2$, where $\gamma_i^0=\frac{P_i|h_{li}|^2}{\sigma^2_{wi}}$ is the average SNR of the $i$-th link. Applying the transformation of random $\gamma_{i} = \gamma_i^0 |h_{i}|^2$ and representing the  incomplete Gamma function using Meijer's G-function,  the PDF  of the SNR for the $i$-th hop can be expressed as
\begin{eqnarray} \label{eq:pdf_hpf_Meijer}
	f_{\gamma_i}(\gamma) = \frac{A_i {\gamma_i^0}^{-\frac{\phi_i}{2}} \gamma^{\frac{\phi_i}{2}-1}}{2} G_{1,2}^{2,0} \Bigg( \frac{C_i {\gamma}^{\frac{\alpha_i}{2}}}{{\gamma_i^0}^{\frac{\alpha_i}{2}}} \Bigg| \begin{matrix} 1 \\ B_i, 0 \end{matrix} \Bigg) 
\end{eqnarray}

We adopt both CA and FG AF relaying protocols for each hop of the multihop-assisted backhaul link. The gain of the CA relay can be computed using the inverse of the channel of the previous hop. Thus, using $G_i^2= \frac{1}{|h_i|^2}$ in  \eqref{eq:multihop_rcvd_signal_eqn}, the end-to-end SNR with the CA relaying in each hop is given by \cite{Hasna_2003_multihop_nakagami_letter}:
\begin{eqnarray} \label{eq:snr_csi_exact}
	\gamma_{N}^{\rm CA} = \Bigg(\sum_{i=1}^{N} \frac{1}{\gamma_{i}}\Bigg)^{-1}
\end{eqnarray}
Note that \eqref{eq:snr_csi_exact} becomes an upper bound if the gain  of the relay also includes the effect of additive noise.
Further, using $G_i= \sqrt{\frac{1}{\psi_i \sigma^2_{wi}}}$ in \eqref{eq:multihop_rcvd_signal_eqn}, where $\psi_i$ is a constant, we can obtain the  end-to-end SNR of  the FG relaying \cite{Hasna_2003_multihop_nakagami_letter}: 
\begin{eqnarray} \label{eq:snr_cg_exact}
	\gamma_{N}^{\rm FG} = \Bigg(\sum_{i=1}^{N} \prod_{j=1}^{i} \frac{\psi_j-1}{\gamma_{j}}\Bigg)^{-1}
\end{eqnarray}
To this end, we interface the multihop backhaul with a single access link by applying DF and FG relaying protocols for uplink and downlink transmissions, respectively. We take into account the shadowing effect along with the channel fading and pointing errors for the THz access link since the transmitted signal may be obstructed by the objects present in a typical access network scenario. We denote the SNR of the access link as $\gamma_{A}= \gamma_A^0 |h_{A}|^2$, where $h_A=  h_{k} h_{p_A}$ and $\gamma_A^0$ is the average SNR of the shadowed access link. Here, $h_{p_A}$ models the pointing error (with parameters $S_A$ and $\phi_A$) and $h_{k}$ models the channel coefficient of the access link. We consider the generalized-$K$ distribution to model the instantaneous power $|h_{k}|^2$ (which includes the short term fading and random path gain) of the shadowed access link \cite{Shankar_2004}:
\begin{eqnarray} \label{eq:pdf_hk}
	&{f_{|h_{k}|^{2} }}(x) = \frac{2b^{m_{A}}}{\Gamma (m_{\Omega}) \Gamma (m_{g}) }x^{\frac{m_{A}}{2}-1}K_{m_{M}}(2b\sqrt{x})
\end{eqnarray}	
where $m_\Omega = \frac{1}{{\rm exp}(\sigma_{n}^2)-1}$, $\sigma_{n}$$>$$0$ denotes the severity of shadowing with factor $\sigma_{\rm dB} =8.686\sigma_{n} $, $m_g$$>$$0$ is frequency dependent parameter characterizing the small-scale fading, $m_A = m_\Omega+m_g$ and $m_M = m_\Omega-m_g$ are the distribution shape parameters of $K_{\nu}(.)$, which is modified Bessel function of the second kind of $\nu$-th order, and $\Gamma(.)$ denotes the gamma function. Denoting $ k_A(f) $ as the absorption coefficient, $d_A$ as the distance of the access link, $G_A$ as the antenna gain, $P_A$ as the transmit power, $\eta$ as the path loss coefficient, $c$ as the speed of light, and $f$ as the carrier frequency, we define $b$ =$ \sqrt{\frac{m_\Omega m_g}{G_A^2 P^{}_{A} \varphi {\rm exp}(\sigma_n^2/2)} \big(\frac{c}{4 \pi  d_A f}\big)^\eta }$, where $\varphi = {\exp}(k_A(f)d_A)$ is the path loss due  the molecular absorption of the signal transmission in the THz band.

\section{Multihop Terahertz Transmission for Backhaul Link}
In this section, we derive analytical expressions for the PDF and CDF of the end-to-end SNR of an $N$-hop  THz wireless system. First, we develop statistical results of the CA-multihop system in terms of multivariate Fox's H-function. Next, we use Mellin transform to derive the exact PDF and CDF of the product of $N$ i.ni.d. random variables distributed according to \eqref{eq:pdf_hpf_Meijer} and raised to a rational power. Finally, we present the outage probability and average BER performance for CA-multihop and FG-multihop links. 

\subsection{Statistical Results}

In the following theorem, we present PDF and CDF of the resultant SNR $\gamma_{N}^{\rm CA}$ (as given in \eqref{eq:snr_csi_exact}) for the CA-multihop system. We use the moment generating function (MGF) for the reciprocal of SNR in \eqref{eq:snr_csi_exact}  to find the CDF  \cite{Hasna_2003_multihop_nakagami_letter}:
\begin{eqnarray} \label{eq:cdf_CSI_exact_main_eqn}
F_{\gamma_N}^{\rm CA} (\gamma) = 1- \mathcal{L}^{-1}\bigg(\frac{\mathcal{M}_{\frac{1}{\gamma_N^{\rm CA}}}(s)}{s}\bigg)\bigg|_{s=1/\gamma_{}}	
\end{eqnarray}
where $\mathcal{L}^{-1}(\cdot)$ denotes the inverse Laplace transform and $s$ is the complex-valued variable. The MGF of sum of $N$ independent random variables $\frac{1}{\gamma_N^{\rm CA}}$ of \eqref{eq:snr_csi_exact} is  defined as
\begin{eqnarray} \label{eq:moment_mh_eqn}
\mathcal{M}_{\frac{1}{\gamma_N^{\rm CA}}}(s) = \prod_{i=1}^{N}\mathcal{M}_{\frac{1}{\gamma_i}}(s) 
\end{eqnarray}
where the MGF of $\frac{1}{\gamma_i}$ is given by
\begin{eqnarray} \label{eq:moment_mh_eqn2}
\mathcal{M}_{\frac{1}{\gamma_i}}(s)=\int_{0}^{\infty} e^{-s/\gamma_{i}} f_{\gamma_{i}}(\gamma) d\gamma 
\end{eqnarray} 
\begin{my_theorem} \label{th:pdf_cdf_mh_exact}
If $\gamma_{i}$ is the SNR of the $i$-th hop distributed according to \eqref {eq:pdf_hpf_Meijer}, then CDF and PDF of the SNR for $N$-hop CA relaying in \eqref{eq:snr_csi_exact} are given as

\begin{eqnarray} \label{eq:cdf_CSI_mh}
F_{\gamma_N}^{\rm CA}(\gamma)=1-\prod_{i=1}^{N}\frac{A_i{\gamma_i^0}^{-\frac{\phi_i}{2}}}{2}{\gamma}^{\big(\sum_{i=1}^{N}\frac{\phi_i}{2}\big)}H_{1,0;1,3;...;1,3}^{0,0;3,0;...;3,0}\left[\begin{matrix}\frac{C_1\gamma^{\frac{\alpha_1}{2}}}{{\gamma_i^0}^{\frac{\alpha_1}{2}} } \\ \vdots \\	\frac{C_N \gamma^{\frac{\alpha_N}{2}}}{{\gamma_i^0}^{\frac{\alpha_N}{2}} } \end{matrix} \Bigg| \begin{matrix} ~\tilde{Q}_1(\gamma)~ \\ ~\tilde{Q}_2(\gamma)~ \end{matrix} \right]   
\end{eqnarray}
where $\tilde{Q}_1(\gamma)$ = $\Big\{\big\{\big(1-\sum_{i=1}^{N} \frac{\phi_i}{2};\frac{\alpha_1}{2},\cdots,\frac{\alpha_N}{2}\big)\big\}; \bigl\{(1,1),\cdots, (1,1)_N\bigr\} \Big\}$ and   $\tilde{Q}_2(\gamma) = \Bigl\{\bigl\{-\bigr\}; \\ \bigl\{(B_1,1), (0,1), \big(\frac{-\phi_1}{2}, \frac{\alpha_1}{2}\big), \cdots,(B_N,1), (0,1), \big(\frac{-\phi_N}{2}, \frac{\alpha_N}{2}\big)\bigr\} \Bigr\}$.

\begin{eqnarray} \label{eq:pdf_CSI_mh}
	 f_{\gamma_N}^{\rm CA}(\gamma) = -\prod_{i=1}^{N} \frac{A_i {\gamma_i^0}^{-\frac{\phi_i}{2}}}{2 } {\gamma}^{\big(\sum_{i=1}^{N}\frac{\phi_i}{2}-1\big)}   H_{2,1;1,3;...;1,3}^{0,1;3,0;...;3,0} \left[ \begin{matrix} \frac{C_1 \gamma^{\frac{\alpha_1}{2}}}{{\gamma_i^0}^{\frac{\alpha_1}{2}} } \\ \vdots \\	\frac{C_N \gamma^{\frac{\alpha_N}{2}}}{{\gamma_i^0}^{\frac{\alpha_N}{2}} } \end{matrix} \Bigg| \begin{matrix} ~Q_1(\gamma)~ \\ ~Q_2(\gamma)~ \end{matrix} \right]   
\end{eqnarray}
where $Q_1(\gamma) = \Big\{\bigl\{\big(\sum_{i=1}^{N}-\frac{\phi_i}{2};\frac{\alpha_1}{2},\cdots, \frac{\alpha_N}{2}\big)\bigr\},\big\{\big(1-\sum_{i=1}^{N} \frac{\phi_i}{2};\frac{\alpha_1}{2},\cdots, \frac{\alpha_N}{2}\big)\big\}; \bigl\{(1,1),\cdots,(1,1)_N\bigr\} \Big\} $ and $Q_2(\gamma) =\Big\{ \bigl\{\big(\sum_{i=1}^{N}1-\frac{\phi_i}{2};\frac{\alpha_1}{2},\cdots, \frac{\alpha_N}{2}\big)\bigr\};  \bigl\{(B_1,1), (0,1), \big(\frac{-\phi_1}{2},\frac{\alpha_1}{2}\big) ,\cdots,(B_N,1), (0,1), \\ \big(\frac{-\phi_N}{2},\frac{\alpha_N}{2}\big)\bigr\} \Big\}$.

\end{my_theorem}
\begin{IEEEproof}	
	The proof is presented in Appendix A.
\end{IEEEproof}

Note that efficient computational programs are available for evaluating the multivariate Fox's H-function of \eqref{eq:pdf_CSI_mh} and \eqref{eq:cdf_CSI_mh} for the CA-multihop system. In the following Corollary, we  provide statistical results of the CA multihop system in terms of simpler functions considering special cases for the THz channel:

\begin{my_corollary}
			\begin{enumerate}[label=(\alph*)]
		\item If $\gamma_{i}$ is the SNR of the $i$-th hop distributed according to \eqref {eq:pdf_hpf_Meijer} with $\alpha_i=2$, $\mu_i=1$ (i.e. Rayleigh fading) and negligible pointing errors $\phi_i\to \infty$, $\forall i$, then the CDF of the SNR for $N$-hop CA relaying  is given as	
				\begin{flalign}
				\label{eq:rayleigh_special}
		F_{\gamma_N}^{\rm CA} (\gamma) = 1- \prod_{i=1}^{N} e^{-\frac{\gamma_{\rm th}}{\gamma_i^0 }}
		\end{flalign}

		\item  If $\gamma_{i}$ is the SNR of the $i$-th hop distributed according to \eqref {eq:pdf_hpf_Meijer} with  $\alpha_i=2$, $\mu_i=2$  (i.e.  Nakagami-$2$ fading) and higher pointing errors $\phi_i=2$, $\forall i$, then the CDF of the SNR for $N$-hop CA relaying  is given as
					\begin{flalign}
						\label{eq:naka_special}
		F_{\gamma_N}^{\rm CA} (\gamma) = 1-  \frac{(\sqrt{\gamma_i^0})^{\frac{1}{2}} \sqrt{S_i}}{2}  \prod_{i=1}^{N} e^{-\frac{2 \gamma_{\rm th}}{\gamma_i^0 S_i^2 }}
		\end{flalign}
		
	\end{enumerate}
\end{my_corollary}

\begin{IEEEproof}
For part (a), we use $\alpha_i=2$, $\mu_i=1$  and  $\phi_i\to \infty$ in \eqref {eq:pdf_hpf_Meijer}, substitute the resultant PDF $f_{\gamma_{i}}(\gamma) \to \frac{1}{\gamma^0_i} e^{-\frac{\gamma_i}{\gamma^0_i}}$ in \eqref{eq:moment_mh_eqn}, and apply the identity \cite[3.324,1]{Gradshteyn} to get $\mathcal{M}_{\frac{1}{\gamma_N^{\rm CA}}}(s) = \prod_{i=1}^N\frac{1}{\gamma^0_i} 2 \sqrt{\gamma_i^0} \sqrt{s}  K\Big[1, \frac{2\sqrt{s}}{\sqrt{\gamma_i^0}}\Big]$, where $K[\cdot, \cdot]$ denotes the modified Bessel function of the second kind. Using  \eqref{eq:cdf_CSI_exact_main_eqn}, and applying  the Laplace inverse transform,  we get the CDF in \eqref{eq:rayleigh_special}. For part (b), we use $\alpha_i=2$, $\mu_i=1$  and  $\phi_i=2$ in \eqref {eq:pdf_hpf_Meijer} to get the resultant PDF $f_{\gamma_{i}}(\gamma) \to \frac{1}{\gamma^0_i} e^{-\frac{\gamma_i}{\gamma^0_i}}$ $f_{\gamma_{i}}(\gamma) = \frac{x}{S_i^2} e^{-\frac{2x^{2}}{S_i^2}}$. Applying the similar procedure for the proof of part (a), we get \eqref{eq:naka_special}.
\end{IEEEproof}

Although multivariate Fox's H-function can be evaluated using computational software \cite{Alhennawi_2016_mv_FoxH}, it is desirable to simplify the analytical expressions in single-variate Fox's H-function for better insights into the system performance. Further, It is known that the FG relaying is simpler than the CA-assisted technique requiring the CSI at each relay. However, analyzing the exact PDF of the end-to-end SNR of the FG-multihop system as depicted in \eqref{eq:snr_cg_exact} is quite complicated. Thus, we use an upper bound on $\gamma_{N}^{\rm FG}$  \cite{Zedini_2015_multihop_AF_FSO_pointing}:
\begin{eqnarray}\label{eq:bound_gen}
\gamma_N^{\rm FG} = \frac{1}{N}\prod_{i=1}^{N} \zeta_i\gamma_{i}^{\frac{l_i}{N}}
\end{eqnarray}
where $l_i=N+1-i$,  $\zeta_i= C_i^{-\frac{(N-i)}{N}}$. 

\begin{my_proposition}\label{prop}
	If $\gamma_i$, $i=1,2, \cdots, N$ are i.ni.d random variables with PDF $f_{\gamma_i}(\gamma)$, then the PDF of $\gamma_N$ in \eqref{eq:bound_gen} is given by
	\begin{eqnarray}\label{eq:gen_prod_pdf_1_mellin2}
	f_{\gamma_N}^{\rm FG}(\gamma) = \frac{1}{\gamma} \frac{1}{2\pi \J} \int_{\mathcal{L}}\frac{1}{N}\prod_{i=1}^N\zeta_i\Aver{\gamma_i^r}\gamma^{-r}    \diff r 
	\end{eqnarray}
	where $\Aver{\gamma_{i}^r}$ denotes the $r$-th moment:
	\end{my_proposition}

\begin{IEEEproof}
We use the Mellin transform pair to develop exact PDF and CDF of \eqref{eq:bound_gen}.  We know that Mellin transform for a function $\phi(x)$ is given as $\phi(t)= \int_{0}^\infty x^{t-1} \phi(x) dx$ with inverse Mellin transform as $\phi(x)= \frac{1}{2\pi \J} \int_{\mathcal{L}} x^{-t}\phi(t)dt$. Expressing the $r$-th moment of $\gamma_N$ in \eqref{eq:bound_gen} as $\Aver{\gamma_N^r}= \int_{0}^\infty \gamma^{r-1} [\gamma f_{\gamma_N}(\gamma)] d\gamma $, we use inverse Mellin  transform to get:
\begin{eqnarray}\label{eq:gen_prod_pdf_1_mellin}
f_{\gamma_N}^{\rm FG}(\gamma) = \frac{1}{\gamma} \frac{1}{2\pi \J} \int_{\mathcal{L}}\gamma^{-r}  \Aver{\gamma_N^r}  \diff r 
\end{eqnarray}
Since random variables $\gamma_i$, $i=1, \cdots, N$ are independent, we substitute $\Aver{\gamma_N^r}= \frac{1}{N}\prod_{i=1}^N\zeta_i\Aver{\gamma_i^r}$ in \eqref{eq:gen_prod_pdf_1_mellin} to get \eqref{eq:gen_prod_pdf_1_mellin2}, where the $r$-th moment is defined as
\begin{eqnarray}\label{eq:rthmoment}
\Aver{\gamma_{i}^r} = \int_{0}^{\infty} \gamma^\frac{rl_i}{N} f_{\gamma_{i}}(\gamma) d\gamma
\end{eqnarray}
\end{IEEEproof}
A straightforward application of the Fox's H-function reveals that  the PDF of the FG-multihop in \eqref{eq:gen_prod_pdf_1_mellin2} can be presented using a single variate Fox's H-function if the  $r$-th moment of the SNR  for the $i$-th hop can be expressed using the ratio of Gamma functions. 

Substituting \eqref{eq:pdf_hpf_Meijer} in  \eqref{eq:rthmoment}  and applying the identity \cite[2.8]{Mathai_2010}, the $r$-th moment of the SNR for the $i$-th hop  can be computed as
\begin{eqnarray}\label{eq:pzaf}
\Aver{\gamma_{i}^{r}} = \frac{A_1 {\gamma_N^0}^{-\frac{\phi}{2}}}{2} \int_{0}^{\infty} {\gamma^{\frac{2rl_i+N\phi}{2N}-1}} G_{1,2}^{2,0} \Bigg(\frac{C_1{\gamma}^{\frac{\alpha}{2}}} {{\gamma_N^0}^{\frac{\alpha}{2}}} \Bigg| \begin{matrix} 1 \\ B_1, 0 \end{matrix} \Bigg) d\gamma
\end{eqnarray}
Substituting $\gamma^{\frac{\alpha}{2}} = t$ with the identity \cite[07.34.21.0009.01]{Meijers}, we solve \eqref{eq:pzaf}:
\begin{eqnarray}\label{eq:pzaf2}
\Aver{\gamma_{i}^{r}} = \frac{A_i {\gamma_N^0}^{-\frac{\phi_i}{2}}}{\alpha_i} \frac{\Gamma\big( {\frac{2r l_i+N\phi_i}{N\alpha_i}}+B_i\big)\Gamma\big({\frac{2rl_i+N\phi_i}{N\alpha_i}}\big)}{\Gamma\big({\frac{2rl_i+N\phi_i}{N\alpha_i}} +1 \big)}\left(C_i{\gamma_N^0}^{-\frac{\alpha_i}{2}}\right)^{-\big(\frac{2r l_i+N\phi_i}{N\alpha_i}\big)}
\end{eqnarray}

In the following Theorem, we present  PDF and CDF of $\gamma^{\rm FG}_N$ to analyze the performance of FG-assisted multihop system:
\begin{my_theorem} \label{th:pdf_cdf_mh_approx}
	If $\gamma_{i}$ is the SNR of the $i$-th hop  distributed according to \eqref {eq:pdf_hpf_Meijer}, then  PDF and CDF of the SNR  for  $N$-hop FG-relaying in \eqref{eq:bound_gen} are bounded as 
	\begin{eqnarray} \label{eq:pdf_multihop}
		f_{\gamma_N}^{\rm FG}(\gamma) = \prod_{i=1}^{N}  \frac{ \zeta_i A_i{C_i^{-\frac{\phi_i}{\alpha_i}}} } {N \alpha_i \gamma} H_{2N,N}^{0,2N} \left[ \prod_{i=1}^{N} \frac{ {\gamma_i^0}^{\frac{ l_i}{N}}}{ C_i^{\frac{2 l_i}{N\alpha_i}} \gamma} \Bigg| \begin{matrix} ~U_1(\gamma)~ \\  ~U_2(\gamma)~ \end{matrix} \right]
	\end{eqnarray}
	
	\begin{eqnarray} \label{eq:cdf_multihop}
		F_{\gamma_N}^{\rm FG}(\gamma) \hspace{-0.5mm}= \hspace{-0.5mm}\prod_{i=1}^{N} \hspace{-0.5mm} \frac{\zeta_i A_i{C_i^{\hspace{-0.5mm}-\frac{\phi_i}{\alpha_i}}} } {N\alpha_i} H_{1+2N,1+N}^{1,2N} \hspace{-0.5mm}\left[\prod_{i=1}^{N} \frac{ {\gamma_i^0}^{\frac{ l_i}{N}}}{ C_i^{\frac{2 l_i}{N\alpha_i}} \gamma} \Bigg| \begin{matrix} U_1(\gamma), \big(1,1\big) \\ \big(0,1\big), U_2(\gamma) \end{matrix} \right]
	\end{eqnarray}
	where $U_1(\gamma) = \bigl\{ \big( 1- \mu_1,\frac{2 l_1}{N\alpha_1} \big),\cdots, \big( 1- \mu_N,\frac{2 l_N}{N\alpha_N} \big), \big( 1-\frac{\phi_1}{\alpha_1}, \frac{2 l_1}{N\alpha_1} \big),\cdots,\big( 1-\frac{\phi_N}{\alpha_N}, \frac{2 l_N}{N\alpha_N} \big) \bigr\}$ and $U_2(\gamma) =  \bigl\{\big(-\frac{\phi_1}{\alpha_1}, \frac{2 l_1}{N\alpha_1} \big), \cdots, \big(-\frac{\phi_N}{\alpha_N}, \frac{2 l_N}{N\alpha_N} \big)\bigr\}$.
\end{my_theorem}

\begin{IEEEproof}	
	The proof is presented in Appendix B.
\end{IEEEproof}
It should be mentioned that the single-variate PDF and CDF derived in Theorem \ref{th:pdf_cdf_mh_approx} can facilitate in developing more insightful analytical expressions for the multihop system albeit with an upper bound performance. 

Further, we can use Proposition \ref{prop} to develop the PDF of the FG-multihop transmission for other wireless systems since the $r$-th moment of the SNR for the $i$-th hop can be expressed using the ratio of Gamma functions for various fading channels of interest. In the following, we present the $r$-th moment of the SNR for some special cases for the THz wireless system:

\begin{my_corollary}
\begin{enumerate}[label=(\alph*)]
		\item If $\gamma_{i}$ is the SNR of the $i$-th hop distributed according to \eqref {eq:pdf_hpf_Meijer} with  $\alpha_i=2$, $\mu=1$  (i.e. Rayleigh) with negligible  pointing errors $\phi_i\to \infty $, $\forall i$, then the $r$-th moment of SNR is:
	\begin{eqnarray} \label{eq:rthmoment_nakagami_zaf2}
	\Aver{\gamma_{i}^r} = \frac{1}{\Omega^{2} {\gamma_i^0}} \Gamma\Big(\frac{rl_i}{N}+1\Big) \Bigg(\frac{-1}{\Omega^{2}{\gamma_i^0}}\Bigg)^{\frac{rl_i}{N}+1}
	\end{eqnarray}	
	\item If $\gamma_{i}$ is the SNR of the $i$-th hop distributed according to \eqref {eq:pdf_hpf_Meijer} with  $\alpha_i=2$, $\mu_i=2$  (i.e.  Nakagami-$2$) and higher pointing errors $\phi_i=2$, $\forall i$,  then the $r$-th moment of SNR is:
	\begin{eqnarray} \label{eq:rthmoment_nakagami_zaf}
\Aver{\gamma_{i}^r} = \frac{1}{S_i^2 \Omega^{4} {\gamma_i^0}} \Gamma\Big(\frac{rl_i}{N}+1\Big) \Bigg(\frac{2}{S_i^2 \Omega^{4} {\gamma_i^0}}\Bigg)^{\frac{rl_i}{N}+1}
\end{eqnarray}

\end{enumerate}
\end{my_corollary}
\begin{IEEEproof}
The proof is straightforward application of the identity \cite[07.34.21.0009.01]{Meijers}.
\end{IEEEproof}
Note that we can use \eqref{eq:rthmoment_nakagami_zaf2} and  \eqref{eq:rthmoment_nakagami_zaf} in \eqref{eq:gen_prod_pdf_1_mellin2} to develop PDF using single-variate Fox's H-functions for the above special THz systems.

\subsection{Performance Analysis for CA-multihop and FG-multihop Backhaul Transmission}
In this subsection, we present analytical expressions for outage probability and average BER for both CA-multihop and FG-multihop systems.
\subsubsection{Outage Probability}
The outage probability is defined as the probability of instantaneous SNR failing to reach a threshold value. Thus, we can use $\gamma = \gamma_{\rm th}$ in \eqref{eq:cdf_CSI_mh} and \eqref{eq:cdf_multihop} to get the outage probability for CA and FG multihop systems, respectively. 

To get an insight on the outage behavior in high SNR regime, we can use \cite{AboRahama_2018} to compute the residue of  multiple Mellin-Barnes integrals at the dominant pole  in  \eqref{eq:cdf_CSI_mh} for the CA-multihop  and apply series expansion of Fox's H-function  \cite[Th. 1.7]{Kilbas_2004} in  \eqref{eq:cdf_multihop} for an asymptotic analysis for the FG-multihop system.  However, the result of \cite{AboRahama_2018} is not directly applicable for the CDF  form in  \eqref{eq:cdf_CSI_mh}. Thus, we substitute \eqref{eq:pdf_CSI_mh} in $\int_{0}^{\gamma}f_{\gamma}(\gamma) d\gamma$ to get an alternative expression for the CDF of the CA-multihop system:
\begin{eqnarray}\label{eq:alt_cdf_ca} 
	F_{\gamma_N}^{\rm CA}(\gamma) = -\prod_{i=1}^{N} \frac{A_i {\gamma}^{\big(\sum_{i=1}^{N}\frac{\phi_i}{2}\big)}}{2 {\gamma_i^0}^{\frac{\phi_i}{2}}} H_{3,1;1,3;...;1,3}^{0,2;3,0;...;3,0} \left[ \begin{matrix} \frac{C_1 \gamma^{\frac{\alpha_1}{2}}}{{\gamma_i^0}^{\frac{\alpha_1}{2}} } \\ \vdots \\	\frac{C_N \gamma^{\frac{\alpha_N}{2}}}{{\gamma_i^0}^{\frac{\alpha_N}{2}} } \end{matrix} \Bigg| \begin{matrix} ~Q_1(\gamma)~ \\ ~Q_2(\gamma)~ \end{matrix} \right]   
\end{eqnarray}
where $Q_1(\gamma) = \Big\{\bigl\{\big(\sum_{i=1}^{N}-\frac{\phi_i}{2};\frac{\alpha_1}{2},\cdots, \frac{\alpha_N}{2}\big), \big(1-\sum_{i=1}^{N}\frac{\phi_i}{2};\frac{\alpha_1}{2},\cdots, \frac{\alpha_N}{2}\big),\big(1-\sum_{i=1}^{N} \frac{\phi_i}{2};\frac{\alpha_1}{2},\cdots, \frac{\alpha_N}{2}\big)\big\}; \\ \bigl\{(1,1),  \cdots,(1,1)_N\bigr\} \Big\} $ and $Q_2(\gamma) =\Big\{ \bigl\{ \big(\sum_{i=1}^{N}-\frac{\phi_i}{2};\frac{\alpha_1}{2},\cdots, \frac{\alpha_N}{2}\big)\bigr\};  \bigl\{(B_1,1), (0,1), \big(\frac{-\phi_1}{2},\frac{\alpha_1}{2}\big) , \\ \cdots,(B_N,1), (0,1),  \big(\frac{-\phi_N}{2},\frac{\alpha_N}{2}\big)\bigr\} \Big\}$.

Now, we apply  \cite{AboRahama_2018} in \eqref{eq:alt_cdf_ca} to compute the residue at the dominant pole  to get the asymptotic outage probability of the CA-multihop link as
\begin{flalign} \label{eq:asymptotic_cdf_mh_CSI}
	\overline{OP}_{\gamma_N}^{\rm CA^\infty} &= -\prod_{i=1}^{N} \frac{A_i {\gamma_i^0}^{-\frac{\phi_i}{2}} {\gamma}^{\big(\sum_{i=1}^{N}\frac{\phi_i}{2}\big)} }{2 \Gamma\big(1-\sum_{i=1}^{N}\frac{\phi_i}{2}+ \sum_{i=1}^{N}\frac{\alpha_i}{2} g_i\big)} \Bigg[ \frac{\sum_{i=1}^{N} \prod_{j=1,j\neq c_i}^{m} \Gamma(b_{i,j} +B_{i,j}-B_{i,j} g_i)}{\sum_{i=1}^{N}\Gamma(2- g_i)} \nonumber \\ & \times \Bigg(\hspace{-1mm} \sum_{i=1}^{N} \frac{C_i \gamma^{\frac{\alpha_i}{2}} }{{\gamma_i^0}^{\frac{\alpha_i}{2}} }\hspace{-1mm}\Bigg)^{g_i} \Bigg]
\end{flalign}
where $g_i = \min\{\mu_i-\frac{\phi_i}{\alpha_i}, 0, -\frac{\phi_i}{\alpha_i}\}$, $b_{i,j} =  \{\mu_i-\frac{\phi_i}{\alpha_i}, 0, -\frac{\phi_i}{2}\}$, $c_i = \arg\min_{j=1:m}\{\frac{b_{i,j}}{B_{i,j}}\}$ , and  $B_{i,j} = \{1, 1, \frac{\alpha_i}{2}\}$.

To derive the outage probability of the FG-multihop system in high SNR region, we apply the asymptotic expansion of a single variate Fox's H-function \cite[Th. 1.7]{Kilbas_2004}:
\begin{eqnarray} \label{eq:outage_asymptotic_mh_cg_ub}
	&\overline{OP}_{\gamma_N}^{\rm FG^\infty} = \prod_{i=1}^{N} \frac {\zeta_i A_i{C_i^{-\frac{\phi_i}{\alpha_i}}} N} {2N l_i}\Bigg[\frac{\prod_{i=1}^{N}\Gamma(-\frac{N\alpha_i\mu_i}{2l_i})\prod_{i=1}^{N}\Gamma(\frac{\phi_i}{\alpha_i}-\mu_i)} {\prod_{i=1}^{N}\Gamma(1+\frac{N\alpha_i\mu_i}{2l_i})\prod_{i=1}^{N}\Gamma(1+\frac{\phi_i}{\alpha_i}-\mu_i)}\bigg(\hspace{-1mm}\prod_{i=1}^{N}\frac{{\gamma_i^0}^{\frac{l_i}{N}}}{C_i^{\frac{2l_i}{N\alpha_i}}\gamma}\bigg)^{\hspace{-1mm}\frac{-N\alpha_i \mu_i}{2 l_i}} \nonumber \\ &+\frac{\prod_{i=1}^{N}\Gamma(-\frac{N\phi_i}{2l_i})\prod_{i=1}^{N}\Gamma(1+\mu_i -\frac{\phi_i}{\alpha_i})} {\prod_{i=1}^{N} \Gamma\big(1+\frac{N\phi_i}{2l_i} \big)} \bigg(\hspace{-1mm} \prod_{i=1}^{N} \frac{{\gamma_i^0}^{\frac{ l_i}{N}}}{C_i^{\frac{2 l_i}{N\alpha_i}} \gamma} \bigg)^{\hspace{-1mm}\frac{ -N\phi_i}{2 l_i}} \Bigg]
\end{eqnarray}

Using $\gamma_i^{0}=\gamma^{0}, \forall i$, and compiling the exponent of $\gamma^{0}$ in  \eqref{eq:asymptotic_cdf_mh_CSI} and \eqref{eq:outage_asymptotic_mh_cg_ub}, the diversity order of for both  CA and FG multihop systems can be obtained as 
\begin{flalign} \label{eq:diversity_multihop_CA}
M_{\rm out}^{\rm CA}= M^{\rm {FG}}_{\rm out}=\min\Biggl\{\sum_{i=1}^N\frac{\alpha_i\mu_i}{2}, \sum_{i=1}^N\frac{\phi_i}{2}\Biggr\}.
\end{flalign}
The diversity order in \eqref{eq:diversity_multihop_CA} shows that multiple relay units in the backhaul link improve the system performance.
\subsubsection{Average BER}
The average BER of a communication system for a general modulation scheme over fading channel with CDF $F(\gamma)$ is given by \cite{Zedini_2017_ber_formula}: 
\begin{eqnarray} \label{eq:ber}
	\bar{P}_e = \frac{\delta}{2\Gamma(p)} \sum_{n=1}^{k} q^p_n \int_{0}^{\infty} \gamma^{p-1} {e^{{-q \gamma}}} F_{\gamma} (\gamma)   d\gamma
\end{eqnarray}
where the constants $p$, $q$, $\delta$ and $k$ determine the type of modulation scheme. Thus, we substitute \eqref{eq:cdf_CSI_mh} in \eqref{eq:ber} to get 
\begin{flalign} \label{eq:cdf_CSI_int}
	\bar{P}_{e_N}^{\rm CA} &= 1-\prod_{i=1}^{N} \frac{A_i {\gamma_i^0}^{-\frac{\phi_i}{2}}\delta }{4\Gamma(p)}  \sum_{n=1}^{k} q^p_n \Bigg[\Big(\frac{1}{2\pi i}\Big)^N \int_{\mathcal{L}_i}^{} \frac{\prod_{i=1}^{N} \Gamma(B_i-S_i) \Gamma(0-S_i) \Gamma\big(-\frac{\phi_i}{2}-\frac{\alpha_iS_i}{2}\big)}{\prod_{i=1}^{N} \Gamma(1-S_i)} \nonumber \\ & \times \bigg(\prod_{i=1}^{N}\frac{ C_i }{{\gamma_i^0}^{\frac{\alpha_i}{2}}}\bigg)^{S_i}  \frac{1}{\Gamma\big(-\sum_{i=1}^{N}\frac{\phi_i+\alpha_iS_i}{2}+1 \big)} dS_i\Bigg] I_{1}  
\end{flalign}
where $ I_{1} =\int_{0}^{\infty} \gamma^{\big(\sum_{i=1}^{N}\frac{\phi_i+\alpha_iS_i}{2}+p-1\big)} {e^{{-q \gamma}}} d\gamma = q^{-\big(\sum_{i=1}^{N}\frac{\phi_i+\alpha_iS_i}{2}+p\big)} \Gamma\Big( \sum_{i=1}^{N}\frac{\phi_i+\alpha_iS_i}{2}+p \Big) $\cite[3.381/4]{Gradshteyn}. To represent the multiple Barnes integrals of \eqref{eq:cdf_CSI_int} in terms of Fox's H-function, we use $S_i\to -S_i$, and apply the definition of Fox's H-function \cite{Mathai_2010} to get the average BER of the CA-multihop:

\begin{eqnarray} \label{eq:ber_CSI}
	\bar{P}_{e_N}^{\rm CA} = 1-\prod_{i=1}^{N} \frac{A_i {\gamma_i^0}^{-\frac{\phi_i}{2}} \delta}{4\Gamma(p)} \sum_{n=1}^{k} \frac{1}{q^{\big(\sum_{i=1}^{N}\frac{\phi_i}{2}\big)}_n}  H_{1,1;3,1;...;3,1}^{0,1;0,3;...;0,3} \left[ \begin{matrix}  \frac{{\gamma_i^0}^{\frac{\alpha_1}{2}}}{C_1 q^{\big(\frac{\alpha_1S_1}{2}\big)}_n} \\ \vdots \\	\frac{{\gamma_i^0}^{\frac{\alpha_N}{2}} }{C_N q^{\big(\frac{\alpha_NS_N}{2}\big)}_n} \end{matrix} \Bigg| \begin{matrix}   ~T_1(\gamma)~  \\  ~T_2(\gamma)~ \end{matrix} \right]
\end{eqnarray}
where $T_1(\gamma) = \Bigl\{ \big\{\big(1-p+\sum_{i=1}^{N} \frac{\phi_i}{2};\frac{\alpha_1}{2},\cdots,\frac{\alpha_N}{2}\big)\big\}; \bigl\{(1-B_1,1), (1,1), \big(1+\frac{\phi_1}{2}, \frac{\alpha_1}{2}\big),\cdots,(1-B_N,1), (1,1), \big(1+\frac{\phi_N}{2}, \frac{\alpha_N}{2}\big)\bigr\} \Bigr\}$ and $T_2(\gamma) = \Bigl\{ \big\{\big(\sum_{i=1}^{N} \frac{\phi_i}{2};\frac{\alpha_1}{2},\cdots,\frac{\alpha_N}{2}\big)\big\}; \bigl\{(0,1),\cdots,(0,1)_N\bigr\} \Bigr\}$.

To get the average BER for the FG-multihop system, we use \eqref{eq:cdf_multihop} in \eqref{eq:ber} with the Fox's H- definition:
\begin{flalign} \label{eq:ber:fg:zaf}
	\bar{P}_{e_N}^{\rm FG} &=\Bigg[ \prod_{i=1}^{N} \frac {\zeta_i A_i{C_i^{-\frac{\phi_i}{\alpha_i}}} } {N\alpha_i}  \frac{1}{2\pi \J} \int_{\mathcal{L}} \prod_{i=1}^{N} \left({C_i^{\frac{2 l_i}{N\alpha_i}} {\gamma_i^0}^{-\frac{ l_i}{N}}}\right)^{-r}  \frac{ \prod_{i=1}^{N} \Gamma\big( {\frac{2r l_i}{N\alpha_i}} +\frac{B_i\alpha_i+\phi_i}{\alpha_i} \big) \Gamma\big( {\frac{2r l_i}{N\alpha_i}}+\frac{\phi_i}{\alpha_i}\big) }{ \prod_{i=1}^{N} \Gamma\big( {\frac{2r l_i}{N\alpha_i}} +\frac{\phi_i+\alpha_i}{\alpha_i} \big)} \nonumber \\ & \times \frac{ \Gamma(-r)}{\Gamma(1-r)} dr\Bigg]  I_3    
\end{flalign}
where $I_3= \frac{\delta}{2\Gamma(p)} \sum_{n=1}^{k} q^p_n \int_{0}^{\infty} e^{-q_n\gamma} \gamma^{p-r-1} d\gamma = q_n^{-p+r} \Gamma(p-r)$ \cite[3.381/4]{Gradshteyn}. Thus, we use the definition of Fox's H-function \cite{Mathai_2010} to express \eqref{eq:ber:fg:zaf} as
\begin{flalign} \label{eq:ber_mh_csi_ub}
	\bar{P}_{e_N}^{\rm FG} \hspace{-1mm} =\hspace{-1mm} \prod_{i=1}^{N} \frac {\zeta_i A_i{C_i^{-\frac{\phi_i}{\alpha_i}}} } {N\alpha_i 2\Gamma(p)} \hspace{-1mm} \sum_{n=1}^{k} \hspace{-1mm} q_n^{-p} H_{1+2N,2+N}^{2,2N} \Bigg[\prod_{i=1}^{N} \frac{{\gamma_i^0}^{\frac{l_i}{N}}q}{C_i^{\frac{2 l_i}{N\alpha_i}}}\Bigg| \begin{matrix} \prod_{i=1}^{N}\big(1-\mu_i,\frac{2 l_i}{N\alpha_i} \big) ,\prod_{i=1}^{N} \big( 1-\frac{\phi_i}{\alpha_i}, \frac{2 l_i}{N\alpha_i} \big), \big(1,1 \big) \\ \big(p,1\big), \big(0,1 \big),\prod_{i=1}^{N} \big( -\frac{\phi_i}{\alpha_i},\frac{2 l_i}{N\alpha_i} \big) \end{matrix} \Bigg]
\end{flalign}
We can obtain asymptotic expression (not written for brevity) for the average BER in high SNR region for the CA and  FG multihop by applying the similar method used in deriving the asymptotic expressions for the outage probability. 

In what follows, we use the statistical results of Theorem \ref{th:pdf_cdf_mh_exact} and Theorem \ref{th:pdf_cdf_mh_approx} to analyze the performance of a mixed system consisting of multihop backhaul link with a shadowed access link.

\section{ Performance Analysis of Multihop Backhaul Mixed with Access Link}
In this section, we analyze the end-to-end performance of a mixed system when an access link is connected to the backhaul. We consider both uplink (UL) and downlink (DL) scenarios. In the uplink transmission,  the near-optimal DF relaying protocol is used to mix the access link with the CA-assisted $N$-hop backhaul link. The DF relay is more appropriate for the uplink since perfect decoding of the transmitted signal can be achieved in the first hop between the access and backhaul links. However, we use the FG relaying to integrate the backhaul and access links in the downlink transmission since perfect decoding of the signal after many hops in the backhaul may not be possible  Moreover, we consider FG-multihop to realize an all FG relaying protocol for the downlink transmission. It should be mentioned that the additional hop between the backhaul and access network may not be readily included as an $N+1$- hop backhaul transmission since the propagation characteristics of the access network may be different from the backhaul.

In the following, we develop statistical results for the access link in the THz band considering the combined effect of short-term fading, shadowing, and pointing errors:
\begin{my_proposition} \label{preposition:pdf_cdf_gk}
	If $\gamma_A$ is the SNR of shadowed access link distributed according to \eqref{eq:pdf_hk}, then the PDF and CDF of the generalized-$K$ fading combined with pointing errors:
	\begin{eqnarray} \label{eq:pdf_kp_combined}
	f_{\gamma_{A}}(\gamma) = \frac{b^{m_{A}} \gamma^{\frac{m_{A}}{2}-1}  \phi_A (S_A^2)^{{\frac{\phi_A-m_{A}}{2}}+1} }{ {\gamma_A^0}^{\frac {m_{A}}{2}} \Gamma(m_{\Omega}) \Gamma(m_{g})} G_{1,3}^{3,0}\left(\frac{b^2 \gamma}{ S_A^2 {\gamma_A^0}} \Bigg| \begin{matrix} 1+\frac{\phi_A-m_{A}}{2} \\  {m_{M}}, {m_{M}}, \frac{\phi_A-m_{A}}{2} \end{matrix}\right)
	\end{eqnarray}
	\begin{eqnarray} \label{eq:cdf_kp_combined}
	F_{\gamma_{A}}(\gamma) = \frac{b^{m_{A}} \gamma^{\frac{m_{A}}{2}} \phi_A (S_A^2)^{{\frac{\phi_A-m_{A}}{2}}+1} }{{\gamma_A^0}^{\frac {m_{A}}{2}} \Gamma(m_{\Omega}) \Gamma(m_{g})}  G_{2,4}^{3,1}\left(\frac{b^2 \gamma}{ S_A^2 {\gamma_A^0}}  \Bigg| \begin{matrix} 1-\frac{m_{A}}{2}, 1+\frac{\phi_A-m_{A}}{2} \\  {m_{M}}, {m_{M}}, \frac{\phi_A-m_{A}}{2}, -\frac{m_{A}}{2} \end{matrix}\right) 
	\end{eqnarray}
\end{my_proposition}
\begin{IEEEproof}
	The proof is presented in Appendix C.  	
\end{IEEEproof}
In the next two subsections, we use \eqref{eq:pdf_kp_combined} and \eqref{eq:cdf_kp_combined} to analyze the performance of the mixed link for both  uplink and downlink transmissions.

\subsection{Uplink Transmissions}

Assuming perfect decoding of the received signal, the resultant SNR of the mixed multihop system for the uplink transmission is given by
\begin{eqnarray} \label{eq:df_comb}
	\gamma=  \min\{\gamma_N^{\rm CA}, \gamma_A\} 
\end{eqnarray}

In the following subsections, we derive exact analytical expressions of the uplink transmissions. As such, we use results of Theorem \ref{th:pdf_cdf_mh_exact} and Proposition \ref{preposition:pdf_cdf_gk} to analyze the system performance.
\subsubsection{Outage Probability}
We use \eqref{eq:cdf_CSI_mh} and \eqref{eq:cdf_kp_combined} at $\gamma = \gamma_{\rm th}$ to get the outage probability of the uplink transmission as \cite{papoulis_2002}  
\begin{eqnarray}\label{eq:df_gen_zaf}
	\overline{OP}^{\rm UL}=  F_{\gamma_N}^{\rm CA}(\gamma_{\rm th})  + F_{\gamma_A}(\gamma_{\rm th}) -F_{\gamma_N}^{\rm CA}(\gamma_{\rm th})F_{\gamma_A}(\gamma_{\rm th})
\end{eqnarray}

To derive the diversity order of the system, asymptotic expressions of the outage probability in high SNR regime for both $F_{\gamma_N}^{\rm CA}(\gamma_{\rm th})$ (as derived in \eqref{eq:asymptotic_cdf_mh_CSI}) and $ F_{\gamma_A}(\gamma_{\rm th})$ is required. To get an asymptotic outage probability of the shadowed access link, we use the asymptotic series expansion of the Meijer's G-function at $\gamma^{0}_A\to 0$ \cite[07.34.06.0006.01]{Meijers} in \eqref{eq:cdf_kp_combined}: 
\begin{eqnarray} \label{eq:asymptotic_outage_kp}
	F_{\gamma_A}^\infty(\gamma) = \frac{b^{m_{A}} \phi_A {\gamma}^{-\frac{m_{A}}{4}} (S_A^2)^{{\frac{\phi_A-m_{A}}{2}}+1}}{ {\gamma_A^0}^{\frac{m_{A}}{2}} \Gamma(m_{\Omega}) \Gamma(m_{g})}  \sum_{k=1}^{m} \frac{\prod_{j=1,j\neq k}^{m} \Gamma(b_j-b_k) \prod_{j=1}^{n} \Gamma(1-a_j+b_k) }{\prod_{j=n+1}^{p} \Gamma(a_j-b_k) \prod_{j=m+1}^{q} \Gamma(1-b_j+b_k) } \Big(\frac{b^2 {\gamma} }{ {\gamma_A^0}}\Big)^{b_k}
\end{eqnarray}
where $b_j = b_k = \big\{ \underbrace{m_{M}, m_{M}, \frac{\phi_A-m_{A}}{2}}_{m}, \underbrace{-\frac{m_{A}}{2}}_{q-m} \big\} $ and $a_j = \big\{ \underbrace{ 1-\frac{m_{A}}{2}}_{n},  \underbrace{1+\frac{\phi_A-m_{A}}{2}}_{p-n} \big\} $. Thus, the diversity order for the access link is $b_k+\frac{m_A}{2}$, where $k=1,\cdots, m$. Using \eqref{eq:asymptotic_cdf_mh_CSI} and \eqref{eq:asymptotic_outage_kp} in \eqref{eq:df_gen_zaf} and  applying $\gamma_i^{0}=\gamma^{0}=\gamma_A^0, \forall i$, the diversity order of the uplink transmissions can be expressed as
\begin{eqnarray}  \label{diversity_order_outage_mixed_csi_exact} 
	{M_{\rm out}^{\rm UL} = \min\Biggl\{ \sum_{i=1}^{N} \frac{\alpha_i\mu_i}{2}, \sum_{i=1}^{N} \frac{\phi_i}{2}, \frac{m_A}{2}+m_M, \frac{\phi_A}{2}  \Biggr\}}
\end{eqnarray}
The diversity order in \eqref{diversity_order_outage_mixed_csi_exact} shows that multiple relay units in the backhaul link improve the system performance. Thus, the asymptotic performance becomes limited to short-term fading and shadowing of the access link when sufficient higher $N$ and higher beam-width for signal transmission in the access link is employed.
\subsubsection{Average BER}
The average BER of the DF relaying system with gray coding be expressed  as \cite{Tsiftsis_2006_multihop_AF_FSO}:
\begin{eqnarray} \label{eq:ber_csi_df_dual_hop_eqn}
	\bar{P}_e^{\rm UL} = \bar{P}_{e_N}^{\rm CA} + \bar{P}_{e_A} - \bar{P}_{e_N}^{\rm CA}\bar{P}_{e_A}
\end{eqnarray}
where $\bar{P}_{e_N}^{\rm CA}$ (already derived in \eqref{eq:ber_CSI}) and $\bar{P}_{e_A}$ are the average BER of backhaul and shadowed access link, respectively. To get $\bar{P}_{eA}$, we use \eqref{eq:cdf_kp_combined} in \eqref{eq:ber} and apply the definition of the Meijer's G-function:
\begin{flalign} \label{eq:cdf_kp_combined_int2}
	\bar{P}_{e_A} =& \Bigg[\frac{b^{m_{A}} \phi_A (S_A^2)^{{\frac{\phi_A-m_{A}}{2}}+1}}{2\Gamma(p) {\gamma_A^0}^{\frac{m_{A}}{2}} \Gamma(m_{\Omega}) \Gamma(m_{g})} \frac{1}{2\pi i} \int_{\mathcal{L}}^{} \frac{ \Gamma\big({\frac{m_{A}}{2}+s}\big)}{\Gamma\big({1+\frac{m_{A}}{2}+s}\big)} \frac{\Gamma\big(m_{M}-s\big) \Gamma\big(m_{M}-s\big) \Gamma\big({{\frac{\phi_A-m_{A}}{2}}}-s \big)} {\Gamma\big(1+ {{\frac{\phi_A-m_{A}}{2}}}-s \big)} \nonumber \\ \times & \bigg(\frac{b^2}{S_A^2 {\gamma_A^0}}\bigg)^s  \Bigg] I_2
\end{flalign}
where $I_2= \frac{\delta}{2\Gamma(p)} \sum_{n=1}^{k} q^p_n \int_{0}^{\gamma} e^{-q_n\gamma} \gamma^{\frac{m_{A}}{2}+{s}+p-1} d\gamma = q_n^{-\big(\frac{m_{A}}{2}+{s}+p\big)} \Gamma\big(\frac{m_{A}}{2}+{s}+p\big) $. Using $I_2$ and applying the definition of Meijer's G-function in   \eqref{eq:cdf_kp_combined_int2}, the average BER for the access link can be expressed as:
\begin{eqnarray} \label{eq:ber_kp_combined}
	\bar{P}_{e_A} = \frac{b^{m_{A}} \phi_A (S_A^2)^{{\frac{\phi-m_{A}}{2}}+1} \delta}{2\Gamma(p) {\gamma_A^0}^{\frac{m_{A}}{2}} \Gamma(m_{\Omega}) \Gamma(m_{g})} \sum_{n=1}^{k} q_n^{-\frac{m_{A}}{2}}  G_{3,4}^{3,2}\left(\frac{b^2}{q_n S_A^2 {\gamma_A^0}} \Bigg| \begin{matrix} 1-\frac{m_{A}}{2}-p, 1-\frac{m_{A}}{2},  1+\frac{\phi_A-m_{A}}{2} \\ {m_{M}}, {m_{M}}, \frac{\phi_A-m_{A}}{2}, -\frac{m_{A}}{2} \end{matrix}\right) 
\end{eqnarray}
Similar to the outage probability, we can obtain the asymptotic expression of the average BER for the uplink system. 
\subsection{Downlink Transmission}
We apply the FG relaying to mix the FG-multihop and access links such that the end-to-end SNR  becomes
\begin{eqnarray}\label{eq:fg_comb}
	\gamma^{\rm DL}= \frac{\gamma_N^{\rm FG}\gamma_A}{\psi+\gamma_A}
\end{eqnarray}
where $\psi =(\mathbb{E}_{\gamma_N}(1+\gamma)^{-1})^{-1}$  is computed using the semi-blind method of statistical information of the received signal \cite{Hasna_2004_AF}. To analyze the all FG-multihop system, we need to derive the PDF of the resultant SNR, as given in \cite{Hasna_2004_AF}:
\begin{eqnarray} \label{eq:pdf_AF_eqn}
f_\gamma^{\rm DL}(z) = \int_{0}^{\infty} f_{\gamma_N^{\rm FG}} \Big(\frac{z(x+\psi)}{x} \Big) f_{\gamma_A}(x) \frac{x+\psi}{x} dx
\end{eqnarray}
In the following Theorem, we derive analytical expressions of the PDF $f_\gamma^{\rm DL}(z)$ and CDF $F_\gamma^{\rm DL}(z)$ using bivariate Fox's H-function.
\begin{my_theorem} \label{th:pdf_cdf_mixed_af}
	If $\gamma_N^{\rm FG}$ is the SNR of $N$-hop backhaul link and $\gamma_A$ is the SNR of the shadowed access link, then CDF and PDF of the resultant SNR in \eqref{eq:fg_comb}   is given by
	\begin{eqnarray} \label{eq:pdf_CG}
		f_{\gamma}^{\rm DL}(\gamma) = \prod_{i=1}^{N} \frac{\zeta_i A_i C_i^{-\frac{\phi_i}{\alpha_i}} b^{m_A} \phi_A (S_A^2)^{\frac{\phi_A-m_A}{2}+1} \psi^{\frac{m_A}{2}} }{2N\alpha_i {\gamma_A^0}^{\frac {m_{A}}{2}} \Gamma(m_{\Omega}) \Gamma(m_g)\gamma} H_{1,0:2N,1+N:1,4}^{0,1:0,2N:4,0} \Bigg[\prod_{i=1}^{N} \frac{ {\gamma_i^0}^{\frac{ l_i}{N}}} { C_i^{\frac{2 l_i}{N\alpha_i}}  \gamma}, \frac{b^2 \psi}{S_A^2 {\gamma_A^0}} \Bigg| 	\begin{matrix} ~V_1(\gamma)~ \\ ~V_2(\gamma)~ \end{matrix} \Bigg]
	\end{eqnarray}
	where $V_1(\gamma) = \Big\{ \big\{\big(1-\frac{m_A}{2}, 1,1\big) \big\}; \big\{\big(1- \mu_1, \frac{2 l_1}{N\alpha_1} \big),\cdots,\big(1- \mu_N, \frac{2 l_N}{N\alpha_N} \big),\big( 1-\frac{\phi_1}{\alpha_1}, \frac{2 l_1}{N\alpha_1} \big),\cdots,\big( 1-\frac{\phi_N}{\alpha_N}, \frac{2 l_N}{N\alpha_N} \big)  \big\}; \big\{1+\frac{\phi_A-m_A}{2},1  \big\} \Big\}$ and $V_2(\gamma) = \Big\{ \big\{-\big\}; \big\{\big(-\frac{\phi_1}{\alpha_1}, \frac{2 l_1}{N\alpha_1} \big),\cdots\big(-\frac{\phi_N}{\alpha_N}, \frac{2 l_N}{N\alpha_N} \big), \big(1,1\big) \big\}; \\ \big\{ \big(m_M,1\big), \big(m_M,1\big), \big(\frac{\phi_A-m_A}{2},1\big), \big(-\frac{m_A}{2},{1}\big) \big\} \Big\}$.
	\begin{flalign} \label{eq:cdf_CG}
		F_{\gamma}^{\rm DL}(\gamma) \hspace{-1mm} = \hspace{-1mm} \prod_{i=1}^{N} \frac{\zeta_i A_i C_i^{-\frac{\phi_i}{\alpha_i}}  b^{m_A} \phi_A   (S_A^2)^{\frac{\phi_A-m_A}{2}+1} \psi^{\frac{m_A}{2}} }{2N\alpha_i {\gamma_A^0}^{\frac {m_{A}}{2}} \Gamma(m_{\Omega}) \Gamma(m_g) }   H_{1,0:1+2N,2+N:1,4}^{0,1:1,2N:4,0} \left[\prod_{i=1}^{N} \frac{  {\gamma_i^0}^{\frac{ l_i}{N}}} {C_i^{\frac{2 l_i}{N\alpha_i}} \gamma_{\rm th}}, \frac{b^2 \psi}{S_A^2 {\gamma_A^0}}  \Bigg|  \begin{matrix} ~\tilde{V}_1(\gamma)~ \\ ~\tilde{V}_2(\gamma)~ \end{matrix} \right]
	\end{flalign}
	where $\tilde{V}_1(\gamma) = \Big\{ \big\{\big(1-\frac{m_A}{2}, 1,1\big) \big\}; \big\{\big(1- \mu_1, \frac{2 l_1}{N\alpha_1} \big), \cdots, \big(1- \mu_N, \frac{2 l_N}{N\alpha_N} \big),\big( 1-\frac{\phi_1}{\alpha_1}, \frac{2 l_1}{N\alpha_1} \big),\cdots,\big( 1-\frac{\phi_N}{\alpha_N}, \frac{2 l_N}{N\alpha_N} \big), \big(1,1\big)  \big\};  \big\{1+\frac{\phi_A-m_A}{2},1  \big\} \Big\}$ and $\tilde{V}_2(\gamma) = \Big\{ \big\{-\big\}; \big\{ \big(0,1\big), \big(-\frac{\phi_1}{\alpha_1}, \frac{2 l_1}{N\alpha_1} \big),  \cdots\big(-\frac{\phi_N}{\alpha_N}, \frac{2 l_N}{N\alpha_N} \big),  \\ \big(1,1\big) \big\}; \big\{ \big(m_M,1\big), \big(m_M,1\big), \big(\frac{\phi_A-m_A}{2},1\big),  \big(-\frac{m_A}{2},{1}\big) \big\} \Big\}$.	
\end{my_theorem}

\begin{IEEEproof}
	The proof is presented in Appendix C. 
\end{IEEEproof}

In what follows next, we use the results of Theorem \ref{th:pdf_cdf_mixed_af} to analyze the system performance for the mixed THz transmissions.

\subsubsection{Outage Probability}
We derive an exact expression of the outage probability by evaluating the CDF in \eqref{eq:cdf_CG} at $\gamma=\gamma_{\rm th}$. We can also develop asymptotic analysis of the outage probability in high SNR regime. However,  the asymptotic expansion of the multivariate Fox's H-function at the dominant pole \cite{AboRahama_2018} can not be readily applied to the bivariate Fox's H-function in  \eqref{eq:cdf_CG} since the average SNR term appears in both numerator and denominator of the contour integrals.   Thus, we expand the Mellin-Barnes integrals of the bivariate Fox's H-function and  adopt the asymptotic expansion of the single-variate Fox's H-function \cite[Th. 1.11]{Kilbas_2004} by converting $S_1 \to -S_1$. We compute the  corresponding poles  of the first contour integral  ${\mathcal{L}}_1$:   $S_1 = \{\frac{m_A+2S_2}{2}$, $\sum_{i=1}^{N} \frac{N\alpha_i\mu_i}{2l_i}$, $\sum_{i=1}^{N} \frac{N\phi_i}{2l_i}$\}  and poles of   the second contour integral ${\mathcal{L}}_2$: $S_2 = \{\frac{m_A}{2}$, $m_M$, $\frac{\phi-m_A}{2}$, $-\frac{m_A}{2}\}$. Evaluating the  residues of the contour integrals at these poles with  $\gamma_i^0 =C$ and  $\gamma_A^0=\gamma_i^0 = \gamma^0, \forall i $, we present the asymptotic expression of the outage probability in  \eqref{eq:asymptotic_outage_fg_af}. Combining the dominant terms of $\gamma^0$ in \eqref{eq:asymptotic_outage_fg_af}, we get the diversity order of the system as
\begin{eqnarray}\label{diversity_order_outage_FGFG}
M_{\rm out}^{\rm DL} =  \min \Biggl\{\sum_{i=1}^{N}\frac{\alpha_i\mu_i}{2}, \sum_{i=1}^{N}\frac{\phi_i}{2}, \frac{m_A}{2}+m_M,  \frac{\phi_A}{2} \Biggl\}
\end{eqnarray}

\begin{figure*}\small
	\begin{flalign} \label{eq:asymptotic_outage_fg_af}
	&\overline{OP}_{\gamma_{N}}^{\rm FG^\infty} = \prod_{i=1}^{N} \frac{\zeta_i A_i C_i^{-\frac{\phi_i}{\alpha_i}} b^{m_A} \phi_A (S_A^2)^{\frac{\phi-m_A}{2}+1} \psi^{\frac{m_A}{2}} }{2N\alpha_i \Gamma(m_{\Omega}) \Gamma(m_g) {\gamma_A^0}^{\frac {m_{A}}{2}} }    \Bigg[   \Bigg( \frac{ \Gamma(m_M-\frac{m_A}{2}) \Gamma(m_M-\frac{m_A}{2}) \Gamma(-{m_A}) \prod_{i=1}^{N}\Gamma(\mu_i)} { \big(\frac{\phi_A}{2}-{m_A}\big) \prod_{i=1}^{N} \frac{\phi_i}{\alpha_i}} \nonumber \\ & \times \bigg(\prod_{i=1}^{N} \frac{C_i^{\frac{2 l_i}{N\alpha_i}} \gamma_{\rm th}}{ {\gamma_i^0}^{\frac{ l_i}{N}} }\bigg)^{m_A} \bigg(\frac{b^2 \psi}{S_A^2 \gamma_A^0}\bigg)^{m_A} +     \frac{2\Gamma\big( -\frac{m_A}{2}- {m_M}\big) \prod_{i=1}^{N}\Gamma\big(\mu_i- \frac{l_i(m_A-2m_M)}{N\alpha_i} \big)}{\big(\frac{m_A-2m_M}{2}\big) \big(1+\frac{\phi-m_A}{2}-m_M\big)  \Gamma\big(\frac{2m_M-m_A}{2}\big) \prod_{i=1}^{N} \big(\frac{\phi_i}{\alpha_i}- \frac{l_i(m_A-2m_M)}{N\alpha_i}\big)}  \nonumber \\ & \times \bigg( \prod_{i=1}^{N} \frac{C_i^{\frac{2 l_i}{N\alpha_i}} \gamma_{\rm th}}{ {\gamma_i^0}^{\frac{ l_i}{N}} }\bigg)^{\frac{m_A+2m_M}{2}}  \bigg(\frac{b^2\psi}{S_A^2 \gamma_A^0}\bigg)^{m_M}  +      \frac{\Gamma\big(m_M-\big(\frac{\phi_A-m_A}{2}\big)\big) \Gamma\big(m_M-\big(\frac{\phi_A-m_A}{2}\big)\big) \prod_{i=1}^{N}\Gamma\big(\mu_i- \frac{l_i(2m_A-\phi_A)}{N\alpha_i} \big)}{\big(\frac{2m_A-\phi_A}{2}\big)  \big(-{\frac{\phi_A}{2}}\big) \Gamma\big(\frac{\phi_A-2m_A}{2}\big) \prod_{i=1}^{N}\big(\frac{\phi_i}{\alpha_i}- \frac{l_i(2m_A-\phi_A)}{N\alpha_i} \big)}  \nonumber \\ & \times \bigg(\prod_{i=1}^{N}\frac{C_i^{\frac{2 l_i}{N\alpha_i}} \gamma_{\rm th}}{ {\gamma_i^0}^{\frac{ l_i}{N}} }\bigg)^{\frac{\phi_A}{2}}  \bigg(\frac{b^2\psi}{S_A^2 \gamma_A^0 }\bigg)^{\frac{\phi_A-m_A}{2}}    +      \frac{\Gamma\big(m_M+\frac{m_A}{2}\big) \Gamma\big(m_M+\frac{m_A}{2}\big) \prod_{i=1}^{N}\Gamma\big(\mu_i-\frac{2l_im_A}{N\alpha_i}\big) } { (m_A) \big(\frac{\phi_A}{2}\big)  \Gamma(-m_A) \prod_{i=1}^{N} \big(\frac{\phi_i}{\alpha_i}- \frac{2l_im_A}{N\alpha_i}\big)  }  \bigg(\frac{b^2\psi}{S_A^2 \gamma_A^0}\bigg)^{\frac{-m_A}{2}} \Bigg)    \nonumber \\ &      + \Bigg(\frac{\Gamma\big( -\frac{m_A}{2}- m_M\big) \prod_{i=1}^{N} 2\Gamma\big(\frac{m_A}{2}-\frac{N \alpha_i\mu_i}{2l_i}+m_M\big) }{\prod_{i=1}^{N} \big(\frac{\phi_i}{\alpha_i} - \mu_i\big) \big(\frac{N \alpha_i\mu_i}{2l_i}\big) \Gamma(-\frac{N \alpha_i\mu_i}{2l_i}) \big(\frac{\phi_A-m_A}{2}-m_M\big)}  \bigg(\frac{b^2 \psi}{S_A^2 \gamma_A^0}\bigg)^{m_M}    \nonumber \\ &   +     \frac{\Gamma\big(m_M-\frac{\phi_A-m_A}{2}\big) \Gamma\big(m_M-\frac{\phi_A-m_A}{2}\big) \Gamma\big( - \frac{\phi_A}{2}\big) \prod_{i=1}^{N} \Gamma\big(-\frac{N \alpha_i\mu_i}{2l_i}+\frac{\phi_A}{2}\big) }{\prod_{i=1}^{N} \big(\frac{\phi_i}{\alpha_i} - \mu_i\big) \big(\frac{N \alpha_i\mu_i}{2l_i}\big) \Gamma(-\frac{N \alpha_i\mu_i}{2l_i})} \bigg(\frac{b^2\psi}{S_A^2\gamma_A^0}\bigg)^{\frac{\phi_A-m_A}{2}}   \nonumber \\ &  +      \frac{\Gamma(m_M+\frac{m_A}{2})\Gamma(m_M+\frac{m_A}{2})}{ \frac{\phi_A}{2} \prod_{i=1}^{N} \big(\frac{\phi_i}{\alpha_i} - \mu_i\big) \big(\frac{N \alpha_i\mu_i}{2l_i}\big)} \bigg(\frac{b^2\psi}{S_A^2\gamma_A^0}\bigg)^{\frac{-m_A}{2}} \Bigg)    \bigg( \prod_{i=1}^{N} \frac{C_i^{\frac{2 l_i}{N\alpha_i}} \gamma_{\rm th}}{ {\gamma_i^0}^{\frac{ l_i}{N}} }\bigg)^{\frac{N \alpha_i\mu_i}{2l_i}}    \nonumber \\ &   +      \Bigg(  \frac{\Gamma\big( -\frac{m_A}{2}- m_M\big) \prod_{i=1}^{N} 2  \Gamma\big(\frac{m_A}{2}-\frac{N \phi_i}{2l_i}+m_M\big)\Gamma\big(\mu_i- \frac{ \phi_i}{\alpha_i}\big) }{\big(\frac{\phi_A-m_A}{2}-m_M\big) \prod_{i=1}^{N} \Gamma\big(-\frac{N \phi_i}{2l_i}\big) \big(\frac{N \phi_i}{2l_i}\big) }   \bigg(\frac{b^2 \psi}{S_A^2\gamma_A^0}\bigg)^{m_M}  \nonumber \\ &    +    \frac{ \Gamma\big(m_M-\frac{\phi_A-m_A}{2}\big) \Gamma\big(m_M-\frac{\phi_A-m_A}{2}\big) \Gamma\big(\frac{-\phi_A}{2}\big) \prod_{i=1}^{N} \Gamma\big(-\frac{N \phi_i}{2l_i}+\frac{\phi_A}{2}\big) \Gamma\big(\mu_i- \frac{ \phi_i}{\alpha_i}\big) }{\prod_{i=1}^{N} \Gamma\big(-\frac{N \phi_i}{2l_i}\big)\big(\frac{N \phi_i}{2l_i}\big)}    \bigg(\frac{b^2\psi}{S_A^2\gamma_A^0}\bigg)^{\frac{\phi_A-m_A}{2}}     \nonumber \\ & +   \frac{ \Gamma\big(m_M+\frac{m_A}{2}\big)\Gamma\big(m_M+\frac{m_A}{2}\big) \prod_{i=1}^{N}\Gamma\big(\mu_i- \frac{ \phi_i}{\alpha_i}\big)  }{ \frac{\phi_A}{2} \prod_{i=1}^{N} (\frac{N \phi_i}{2l_i})}  \bigg(\frac{b^2\psi}{S_A^2\gamma_A^0}\bigg)^{\frac{-m_A}{2}}   \Bigg)    \bigg(\prod_{i=1}^{N} \frac{C_i^{\frac{2 l_i}{N\alpha_i}} \gamma_{\rm th}}{ {\gamma_i^0}^{\frac{ l_i}{N}} }\bigg)^{\frac{N \phi_i}{2l_i}}    \Bigg]
	\end{flalign}
	\hrule
\end{figure*}

\subsubsection{Average BER}
Next, we analyze the average BER of the downlink transmission by substituting \eqref{eq:cdf_CG} in \eqref{eq:ber} to get
\begin{flalign} \label{eq:ber_CG_int}
	\bar{P}_e^{\rm DL} =& \Bigg[\prod_{i=1}^{N} \frac{\zeta_i A_i C_i^{-\frac{\phi_i}{\alpha_i}}  b^{m_A} \phi_A (S_A^2)^{\frac{\phi_A-m_A}{2}+1} \psi^{\frac{m_A}{2}}}{4 N\Gamma(p)\alpha_i {\gamma_A^0}^{\frac {m_{A}}{2}} \Gamma(m_{\Omega}) \Gamma(m_g) }  \frac{1}{2\pi i}  \int_{\mathcal{L}_1}^{} \Gamma\bigg(\frac{m_A}{2}+S_1+S_2\bigg)   \nonumber \\ \times &  \frac{\prod_{i=1}^{N}\Gamma\big( \frac{B_i\alpha_i+\phi_i}{\alpha_i}+ \frac{2 l_i}{N\alpha_i} S_1 \big) \Gamma\big(\frac{\phi_i}{\alpha_i} + \frac{2 l_i}{N\alpha_i}S_1\big) } {\prod_{i=1}^{N} \Gamma\big(\frac{\phi_i+\alpha_i}{\alpha_i} + \frac{2 l_i}{N\alpha_i}S_1 \big) \Gamma(S_1) }  \frac{\Gamma(-S_1)}{\Gamma(1-S_1)}  \Bigg( \prod_{i=1}^{N}  \frac{{\gamma_i^0}^{\frac{ l_i}{N}}} {C_i^{\frac{2 l_i}{N\alpha_i}}} \Bigg)^{S_1}  dS_1 \nonumber \\ \times &  \frac{1}{2\pi i} \int_{\mathcal{L}_2}^{} \hspace{-2mm} \frac{\Gamma(m_M-S_2) \Gamma(m_M-S_2) \Gamma\big(\frac{\phi_A-m_A}{2}-S_2 \big) \Gamma\big( -\frac{m_A}{2}- {S_2}\big)}{\Gamma\big(1+\frac{\phi_A-m_A}{2}-S_2  \big)}  \bigg( \frac{b^2 \psi}{S_A^2 {\gamma_A^0}} \bigg)^{S_2} dS_2\Bigg] I_7
\end{flalign}
where $ I_7$ = $ \frac{\delta}{2\Gamma(p)} \sum_{n=1}^{k} q^p_n \int_{0}^{\infty} \gamma^{p-1-S_1} e^{-q_n\gamma} d\gamma$ is solved utilizing \cite[3.381/4]{Gradshteyn} to get $q_n^{S_1-p} \Gamma(p-S_1)$. Substituting $I_7$ in \eqref{eq:ber_CG_int} and applying the definition of Fox's H-function \cite{Mathai_2010}, we get the average BER for downlink transmissions as
\begin{eqnarray} \label{eq:ber_CG}
	\bar{P}_e^{\rm DL} = \prod_{i=1}^{N} \frac{\zeta_i A_i C_i^{-\frac{\phi_i}{\alpha_i}}  b^{m_A} \phi_A   (S_A^2)^{\frac{\phi_A-m_A}{2}+1} \psi^{\frac{m_A}{2}} \delta}{4N\Gamma(p)\alpha_i {\gamma_A^0}^{\frac {m_{A}}{2}} \Gamma(m_{\Omega}) \Gamma(m_g) } \sum_{n=1}^{k} H_{1,0:1+2N,3+N:1,4}^{0,1:2,2N:4,0}  \Bigg[\prod_{i=1}^{N} \frac{ {\gamma_i^0}^{\frac{ l_i}{N}} q_n}{ C_i^{\frac{2 l_i}{N\alpha_i}}}, \frac{b^2 \psi}{S_A^2 {\gamma_A^0}}  \Bigg| \begin{matrix} ~X_1(\gamma)~ \\ ~X_2(\gamma)~ \end{matrix} \Bigg]
\end{eqnarray}
where $X_1(\gamma) = \Big\{ \big\{\big(1-\frac{m_A}{4}, 1,1\big) \big\}; \big\{\big(1- \mu_1, \frac{2 l_1}{N\alpha_1} \big),\cdots,  \big(1- \mu_N, \frac{2 l_N}{N\alpha_N} \big),\big( 1-\frac{\phi_1}{\alpha_1}, \frac{2 l_1}{N\alpha_1} \big),\cdots,\big( 1-\frac{\phi_N}{\alpha_N}, \frac{2 l_N}{N\alpha_N} \big), \big(1,1\big)  \big\};  \big\{1+\frac{\phi_A-m_A}{2},1  \big\} \Big\}$ and $X_2(\gamma) = \Big\{ \big\{-\big\}; \big\{ \big(p,1\big), \big(0,1\big),  \big(-\frac{\phi_1}{\alpha_1}, \frac{2 l_1}{N\alpha_1} \big), \cdots\big(-\frac{\phi_N}{\alpha_N}, \frac{2 l_N}{N\alpha_N} \big), \big(1,1\big) \big\}; \big\{ \big(m_M,1\big), \big(m_M,1\big),   \big(\frac{\phi_A-m_A}{2},1\big), \big(-\frac{m_A}{2},{1}\big) \big\} \Big\}$.

The asymptotic expression of average BER for the downlink transmissions can be obtained by following the similar procedure as adopted for outage probability with the same diversity order expression, as given in \eqref{diversity_order_outage_FGFG}.

\section{Simulation and Numerical Analysis}\label{sec:sim_results}
\begin{table}[tp] 
	\caption{List of Simulation Parameters} 
	\label{tab:simulation_parameters} 
	\centering 
	\begin{tabular}{c c c c} 
		\hline\hline   
		\textbf{Parameter}  & \textbf{Value} &  \textbf{Parameter}  & \textbf{Value} \\ [1ex] 
		\hline  
		THz carrier frequency & $ 300 $ \mbox{GHz} \cite{Sen_2020_Teranova} & $\alpha$ & $1$-$2$ \cite{Papasotiriou2021}\\  
		THz signal bandwidth & $ 10 $ \mbox{GHz} \cite{Sen_2020_Teranova}& $\mu$ & $0.5$-$2$\cite{Papasotiriou2021} \\
		Antenna Gain (Backhaul link)  & $ 33 $ \mbox{dBi}\cite{Rappaport_2019}&  $\Omega$ & $1$ \\
		Antenna Gain (Access link) &  $ 10 $ \mbox{dBi} &  $\sigma_{dB}$ & $2$-$8.6$ \cite{Ye_2021_generalized_k} \\
		Transmit power & $0$-$40$ \mbox{dBm}& $m_\Omega$ & $-$ \\
		Noise PSD & $ -174 $ \mbox{dBm/Hz} \cite{Sen_2020_Teranova}& $m_g$ & $1$-$5$ \\
		Noise figure & $ 5 $ \mbox{dB} \cite{Sen_2020_Teranova}& $\phi_i$ & $37$ \\
		Antenna aperture radius $(r_1)$ & $ 10 $ \mbox{cm}& $\phi_A$ & $2.3-14.5$ \\		
		Jitter standard deviation ($\sigma_s$) & $5$-$20$ \mbox{cm} \cite{Farid2007}& $w_z/r$ & $6$ \cite{Farid2007}\\
		\hline \hline 
	\end{tabular}	
\end{table}

In this section, we demonstrate the performances of multihop-assisted backhaul mixed with the shadowed access link using computer simulations. We validate the derived analytical and asymptotic expressions with Monte-Carlo simulations by numerically evaluating Meijer's G and Fox's H-functions. To compute Meijer's G-function and single-variate Fox's H-function, we use the standard library in MATLAB and Mathematica, and use implementation codes of bivariate Fox's H-function and multivariate Fox's H-function \cite{Alhennawi_2016_mv_FoxH} \cite{Illi_2017}. We consider varying backhaul link distance in the range of $d_{\rm MH}=20$ \mbox{m} to $d_{\rm MH}=100$ \mbox{m} with equidistant relay position and a fixed distance of $d_{A}=20$ \mbox{m} for the access link.  The absorption coefficient $k_i$ and $k_A$ is calculated for both backhaul and access link using relative humidity $50\%$, atmospheric pressure $101325$ \mbox{Pa}, temperature $296^{o}$ \mbox{K} at a carrier frequency $300$ \mbox{GHz} \cite{Boulogeorgos_Error,Pranay_2021_TVT}. We list the parameters for channel fading, AWGN, and pointing errors in Table \ref{tab:simulation_parameters}. As such, we consider various non-linearity parameters $\alpha$ and multipath clustering $\mu$ for the backhaul link with varying shadowing severity ($\sigma_{\rm dB}$) and short term fading ($m_{g}$) in the access link. Unless specified, we adopt higher pointing errors in the access link than the backhaul link, i.e., $\phi_A<\phi_i$, $\forall i$.

\subsection{Multihop Backhaul Transmission}
\begin{figure}[tp]	
	\subfigure[Outage Probability, $\mu=1.2$, $\sigma_{s}=5 \mbox{cm}$.]{\includegraphics[scale= 0.43] {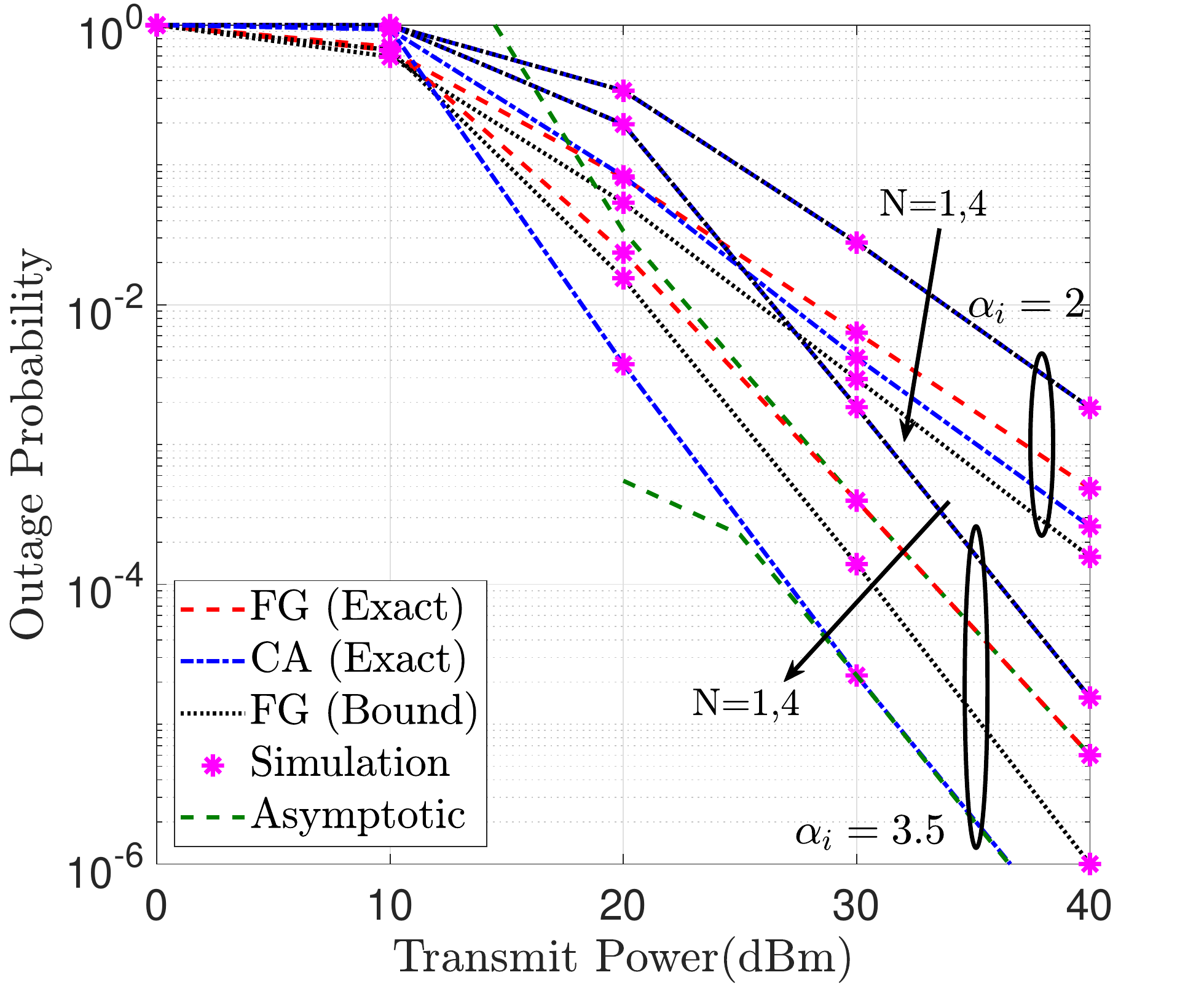}}
	\subfigure[Average BER, $\alpha=1.5$, $\mu=2.5$.] {\includegraphics[scale = 0.43]{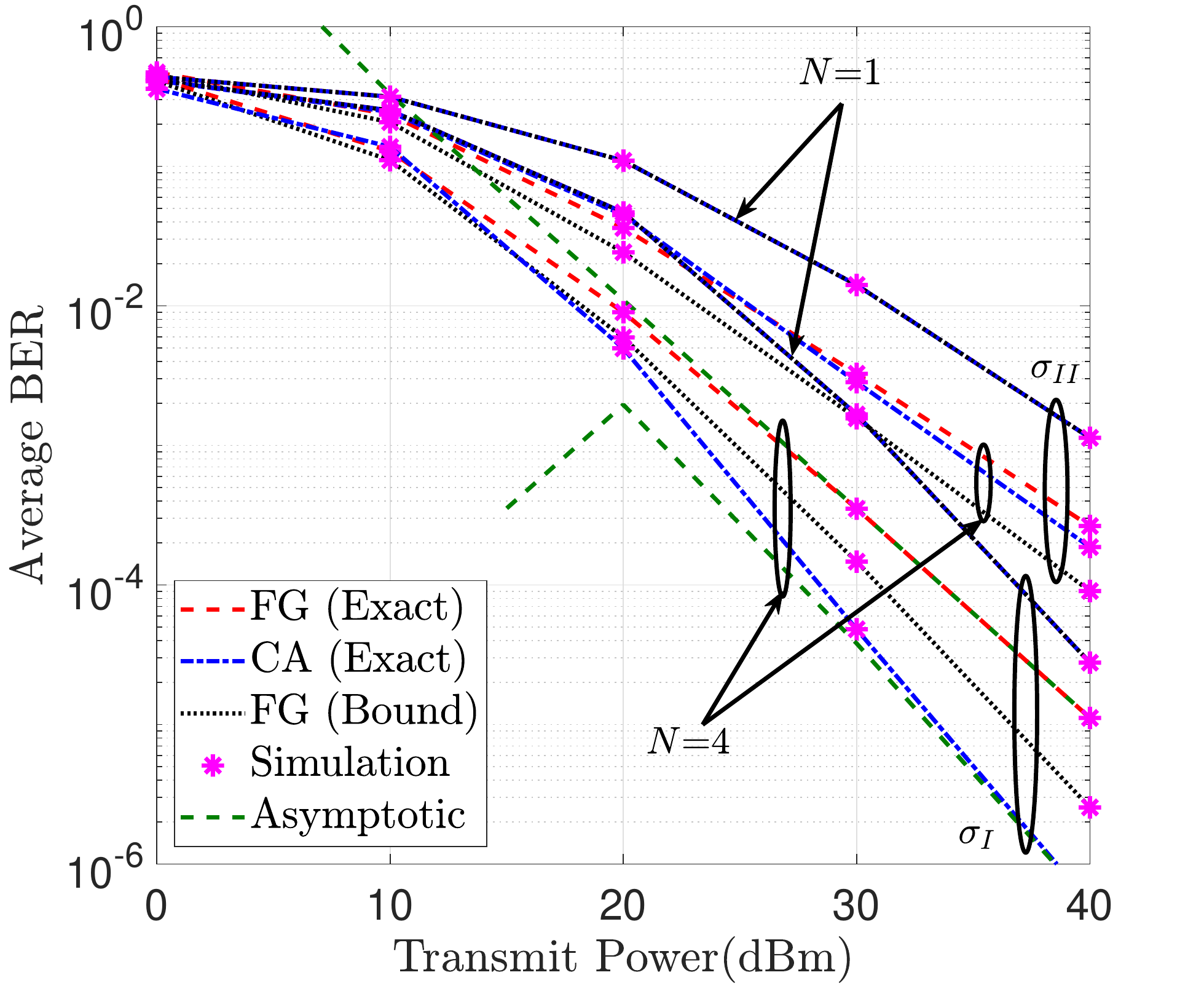}}
	\caption{Performance of multihop THz backhaul link with  $100$ \mbox{m}.}
	\label{fig:n_hop}
\end{figure}

In Fig. \ref{fig:n_hop}, we demonstrate the effect of system and channel parameters on the outage and average BER performance versus transmit power for the backhaul link. We consider i.i.d. channel parameters for each hop, which is a reasonable assumption for short link distances. The outage probability improves with an increase in the fading parameter $\alpha_i$ since the channel becomes more linear, as depicted in  Fig \ref{fig:n_hop}(a). The figure shows that a change in the parameter $\alpha_i=2$ to $\alpha_i=3.5$  requires $15$\mbox{dBm} less transmit power to achieve an outage probability of $5\times 10^{-3}$ with the 4-hop  CA  system. Further, the FG-multihop transmission performs close to the CA-multihop for highly non-linear THz channels. The benefit of multiple relays is also evident from the figure when the direct link ($N=1$) plot is compared with the $4$ hop transmission. It can be seen that the outage performance improves $100$ folds for the 4-hop FG system compared with the direct transmission at an average SNR of $30$ \mbox{dB}.

In Fig. \ref{fig:n_hop}(b), we simulate i.ni.d. pointing errors for each hop to demonstrate the average BER performance of the multihop system. We consider two scenarios of pointing errors: standard deviation of jitter  $\sigma_{I}: \sigma_s=5+i$  \mbox{cm} for $i=1,2,3,4$ hops and $\sigma_{II}:\sigma_s=17+i$ \mbox{cm} for $i=1,2,3,4$ hops. The figure shows a significant degradation in the average BER with an increase in the jitter. The FG-multihop performs close to the CA-multihop when the effect of pointing errors is high. Further, an increase in the number of hops from $N=1$ to $N=4$ reduces the required transmit power by $10$\mbox{dBm} and $5$\mbox{dBm} for the CA and FG multihop systems, respectively.

In the next two subsections, we integrate the backhaul with the access and delve into the inquisition ``how many AF-assisted relays are required in the backhaul to extend the communication range with $N+1$ hops to achieve at least the same performance of the mixed link with the popular dual-hop system ($N=2$) as a reference?" Thus, we demonstrate the impact of the number of hops in the backhaul link on the outage probability and average BER of the mixed link for both uplink and downlink transmission. Unless stated otherwise, we consider the transmission distance of the backhaul link $d_{\rm MG}=20$ \mbox{m} for the $N=2$ scenario and the extended link distance $d_{\rm MH}=50$ \mbox{m} with  multihop relaying at a fixed access link $d_{A}=20$\mbox{m}.

\subsection{Uplink and Downlink Transmission for Mixed Backhaul-Access Link}

\begin{figure}[tp]	
	\centering
	\includegraphics[scale= 0.55]{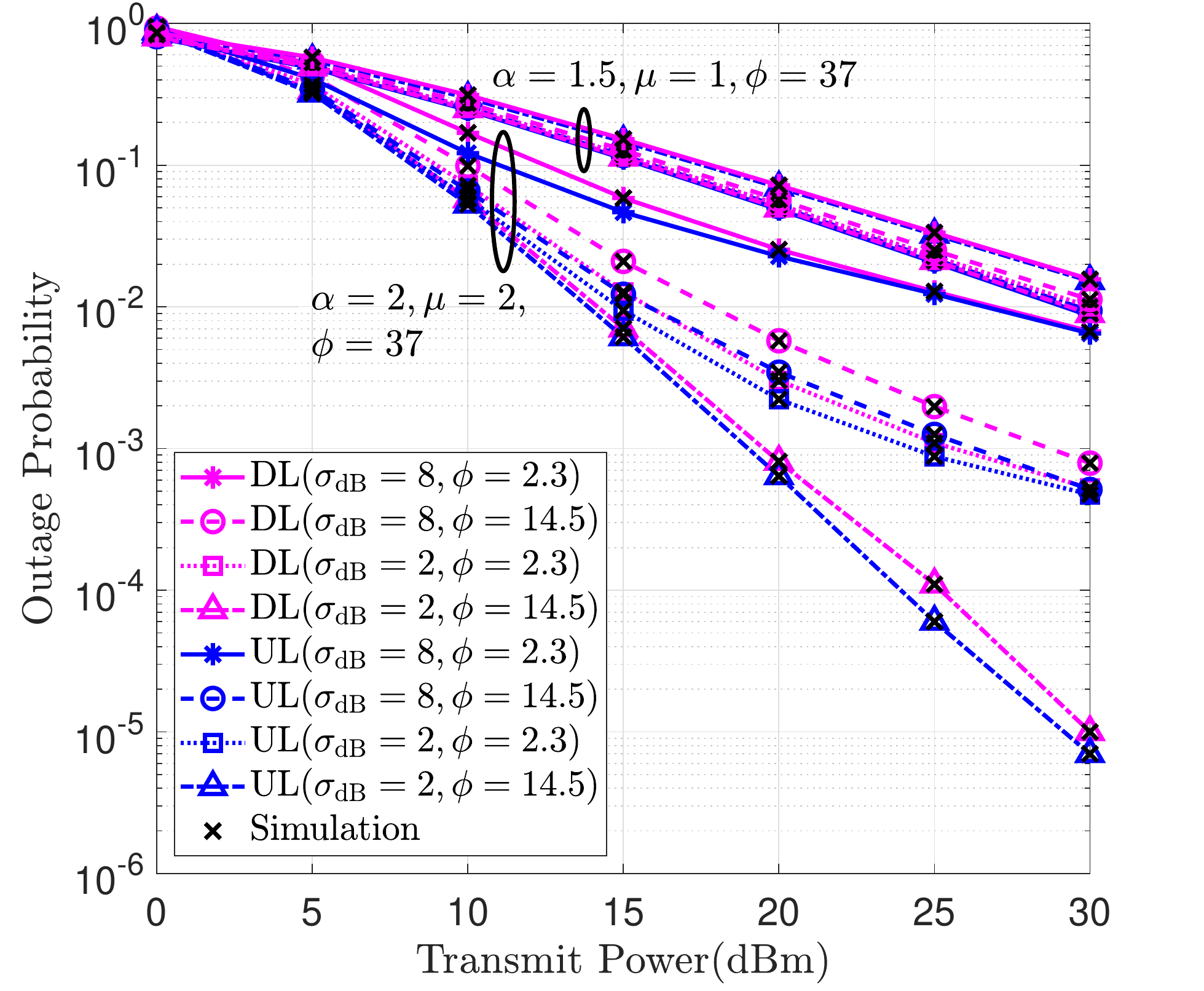}	
	\caption{{\color{blue}Outage probability of the mixed dual-hop relaying system with $m_g=1$, $\phi_i=37$, $d_{\rm MH}=20$ \mbox{m}, and $d_A=20$ \mbox{m}.}}
	\label{fig:outage_dual_hop}
\end{figure}

\begin{figure}[tp]
	\subfigure[Uplink transmission.]{\includegraphics[scale= 0.43] {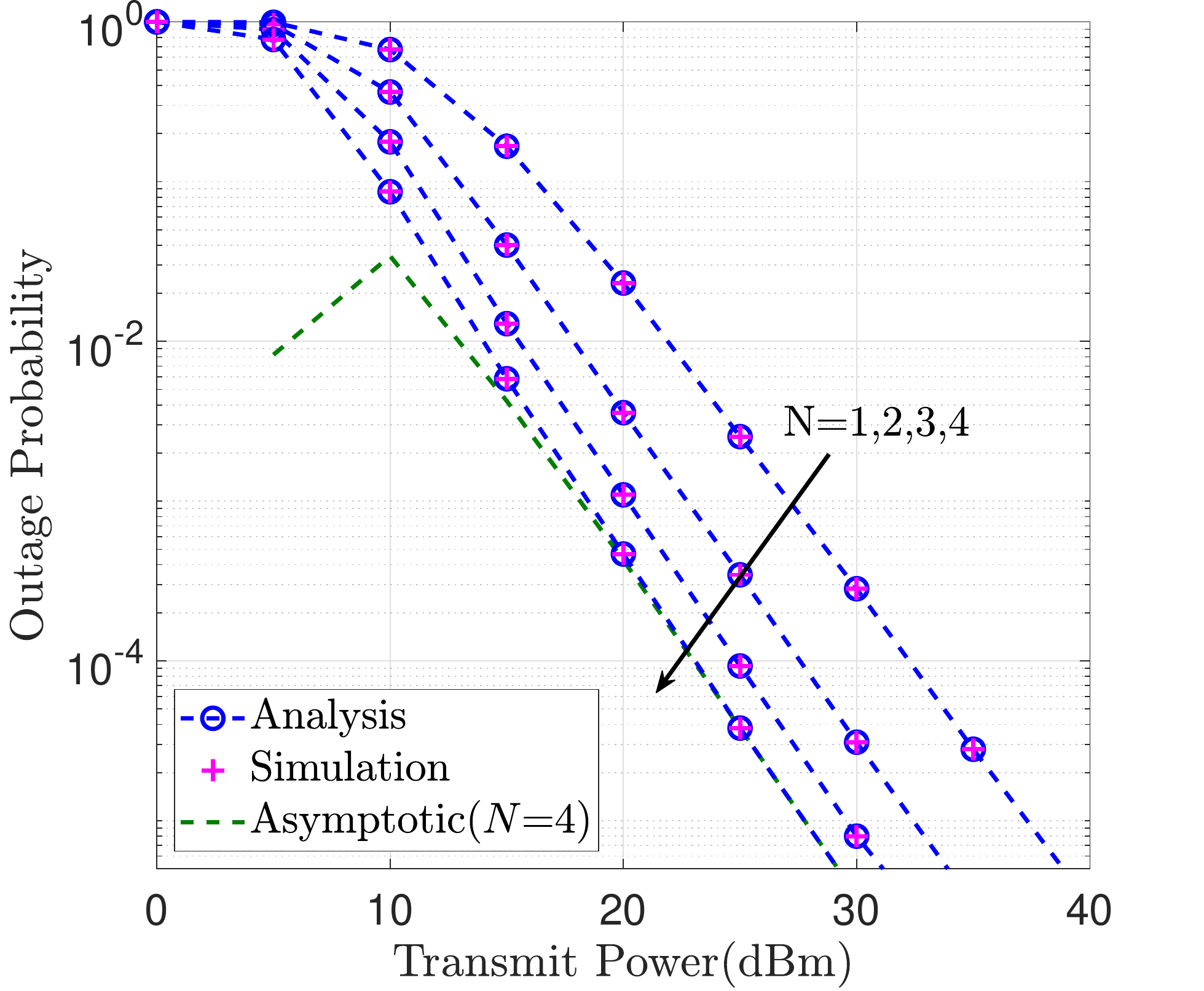}}
	\subfigure[Downlink transmission.]{\includegraphics[scale= 0.43] {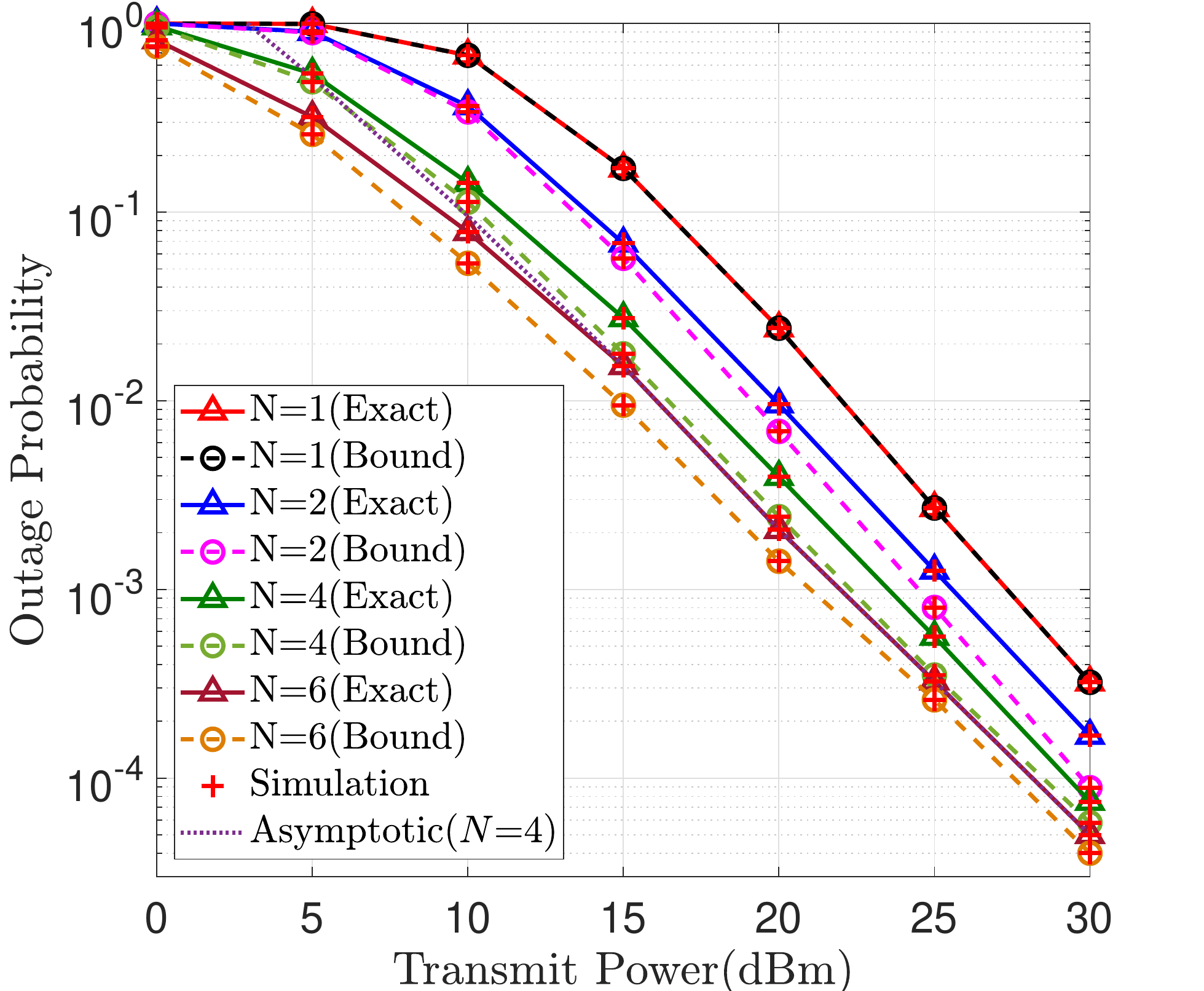}}		
	\caption{Outage probability of the mixed multihop relaying system  with $m_g=1$, $\phi_i=37$, $d_{\rm MH}=50$ \mbox{m} and $d_A=20$ \mbox{m}.}
	\label{fig:outage}
\end{figure}

We plot the outage probability of the mixed relaying system for both uplink and downlink in Fig.~\ref{fig:outage_dual_hop} and \ref{fig:outage}. We demonstrate the impact of channel fading, shadowing, and pointing errors on the performance of the mixed system by considering a single relay (i.e., a typical dual-hop transmission with $N=1$) to integrate the backhaul with the access system in Fig.~\ref{fig:outage_dual_hop}. We vary channel characteristics of the access link at different shadowing factors $\sigma_{\rm dB}=2$ and $\sigma_{\rm dB}=8$ at a short-term fading $m_g=1$ considering both lower ($\phi_A=2.3$) and higher ($\phi_A=14.5$) pointing errors with the backhaul link consisting of two sets $\alpha=2$, $\mu=2$ and $\alpha=1.5$, $\mu=1$ with pointing error parameter $\phi=37$. Note that $\phi=37$, $\phi_A=14.5$, and $2.3$ corresponds to the jitters variance of $\sigma_s^2=5$ \mbox{cm}, $\sigma_s^2=8$ \mbox{cm}, and $20$ \mbox{cm}, respectively \cite{Farid2007}. We fix the link distance of $20$ \mbox {m} for both the hops. It can be seen from Fig.~\ref{fig:outage_dual_hop} that higher values of $\alpha$ and $\mu$ improve the outage probability since an increase in $\alpha$ and $\mu$ reduces the non-linearity effect and improves multipath clustering, respectively.  Further, Fig.~\ref{fig:outage_dual_hop} shows that an increase in the pointing errors (depicted by a decrease of $\phi_A$ from $14.5$ to $2.3$) of the access link decreases the outage performance by a factor $10$ when $\alpha=2$, $\mu=2$, and $\phi=14.5$ at a transmit power of $30$\mbox {dBm}. Moreover, the outage probability is significantly degraded when the shadowing factor is increased from $2$\mbox{dB} (less shadowing) to $8$\mbox{dB} (severe shadowing).

To this end, we choose judicially channel parameters in Fig.~\ref{fig:outage_dual_hop} to demonstrate the diversity order of the system. Note that $m_A= m_\Omega+m_g$ and $m_M=m_\Omega-m_g$, where $m_\Omega= 18.36$ when $\sigma_{\rm dB}=2$ and $m_\Omega= 0.75$ when $\sigma_{\rm dB}=8$. Using $M_{\rm out}=\min \Bigl\{\frac{\alpha\mu}{2}, \frac{\phi}{2}, \frac{m_A}{2}+m_M, \frac{\phi_A}{2} \Bigr\}$, the diversity order is $M_{\rm out}= \frac{\alpha \mu}{2}=0.75$ (when $\alpha=1.5$ and $\mu=1$) since $\frac{\alpha\mu}{2}$ is the minimum among other channel  and pointing error parameters. Thus, the slope of plots for $\alpha=1.5$ and $\mu=1$ in Fig.~\ref{fig:outage_dual_hop} remains independent of other channel parameters i.e., $\sigma_{\rm dB}$,  $\phi_A$, and $m_g$. However, when  $\alpha=2$ and $\mu=2$, the outage-diversity order $M_{\rm out}= 0.62$ depends on the fading parameters of the access link when $\sigma_{\rm dB}= 8$ and becomes  $M_{\rm out}= 1.15$ when $\sigma_{\rm dB}= 2$, which depends on the minimum of pointing errors of the backhaul and shadowed links. Hence, the slope of the plots in Fig.~\ref{fig:outage_dual_hop} confirms our analysis of the diversity order.

We demonstrate the impact of multihop relaying on the outage performance as illustrated in Fig.~\ref{fig:outage}(a) ({\color{blue}for uplink}) and Fig.~\ref{fig:outage}(b) ({\color{blue}for downink}) by considering $\sigma_{dB}=2$, $\phi_A=14.5$, $\alpha=2$, and $\mu=2$. It can be seen that the outage probability of the mixed link significantly improves with an increase in the number of relays employed in the backhaul link. As such the uplink transmission with an $N=4$ relays requires $10$ \mbox{dBm} lesser transmit power to get a typical outage probability of $3\times10^{-3}$, as shown in Fig.~\ref{fig:outage}(a). However, there is a $5$ \mbox{dBm} transmit power saving with the downlink transmission, as depicted in Fig.~\ref{fig:outage}(b) due to the folding of background noise from previous hops. Comparing the two AF relaying schemes, the FG-assisted system requires $6$ hops as compared with only $3$ hops when the CA relaying is employed to achieve a similar outage performance $2\times10^{-4}$ at a transmit power of $25$ \mbox{dBm}. Further, observing the scaling of outage probability with $N$, it can be seen that the cumulative gain of the AF-assisted multihop system (see Fig.~\ref{fig:outage}(a) and  Fig.~\ref{fig:outage}(b)) reduces with an increase in the number of hops advocating the use of a limited number of relays. Comparing the outage probability in  Fig.~\ref{fig:outage_dual_hop} and plot with $N=4$ in Fig.~\ref{fig:outage}(b), it can be seen that the multihop relaying in the backhaul extends the communication range from $d_1=20$ \mbox{m} to $d_1=50$ \mbox{m} with similar outage performance using $3$ relays providing an average of $8$ \mbox{m} range extension with each additional relay. The effect of multihop relaying $N$ on the diversity order can also be observed from the slope of plots, as shown in   Fig.~\ref{fig:outage}(a) and Fig.~\ref{fig:outage}(b).

\begin{figure}[tp]
	\centering	
	\includegraphics[scale = 0.55]{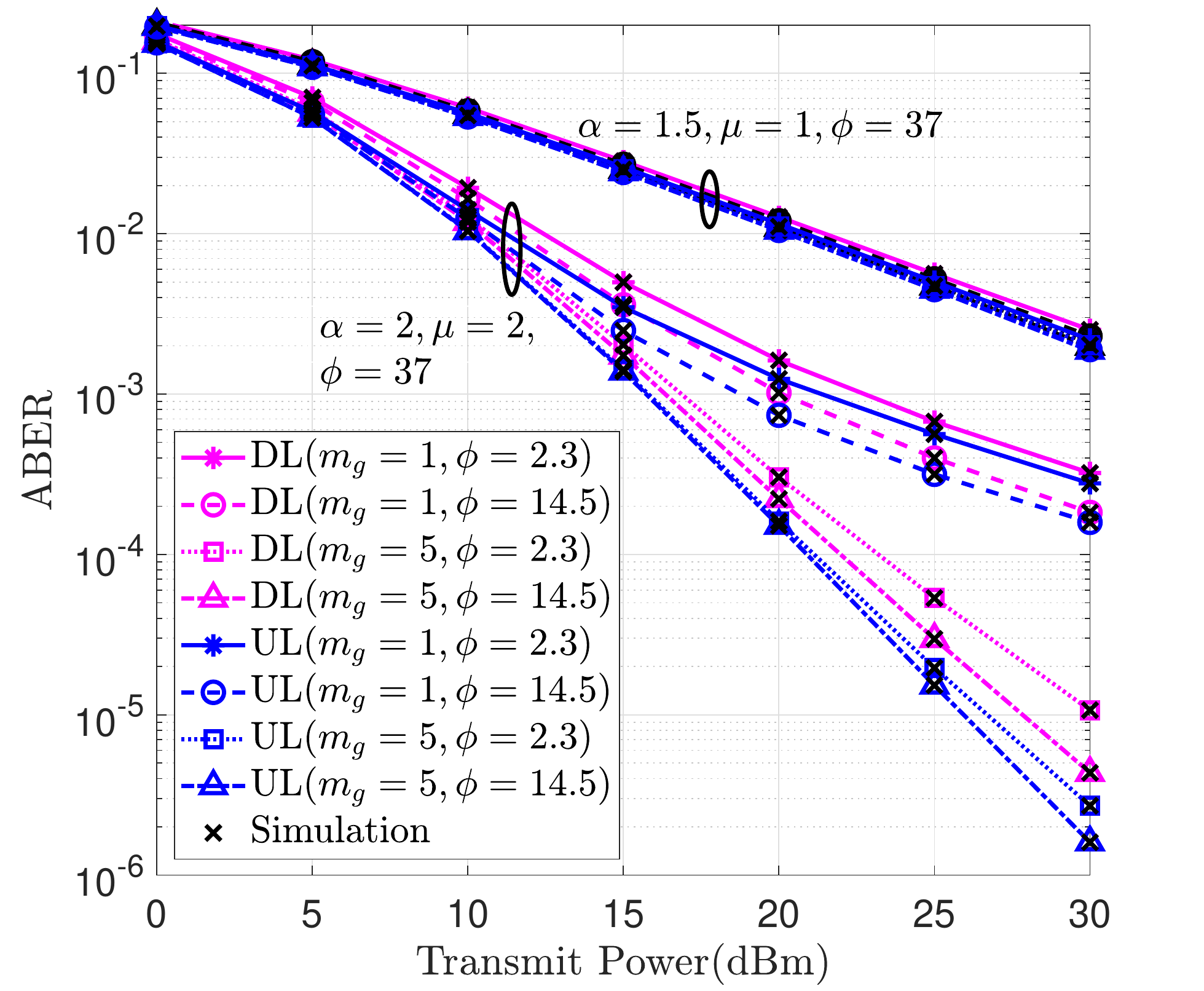}
	\caption{Average BER of the mixed dual-hop relaying system with $\sigma_{dB}=5$, $\phi_i=37$, $d_{\rm MH}=20$ \mbox{m}, and $d_A=20$\mbox{m}.}
	\label{fig:ber_dual_hop}
\end{figure}

\begin{figure}[tp]	
	\subfigure[Uplink transmission.]{\includegraphics[scale = 0.43]{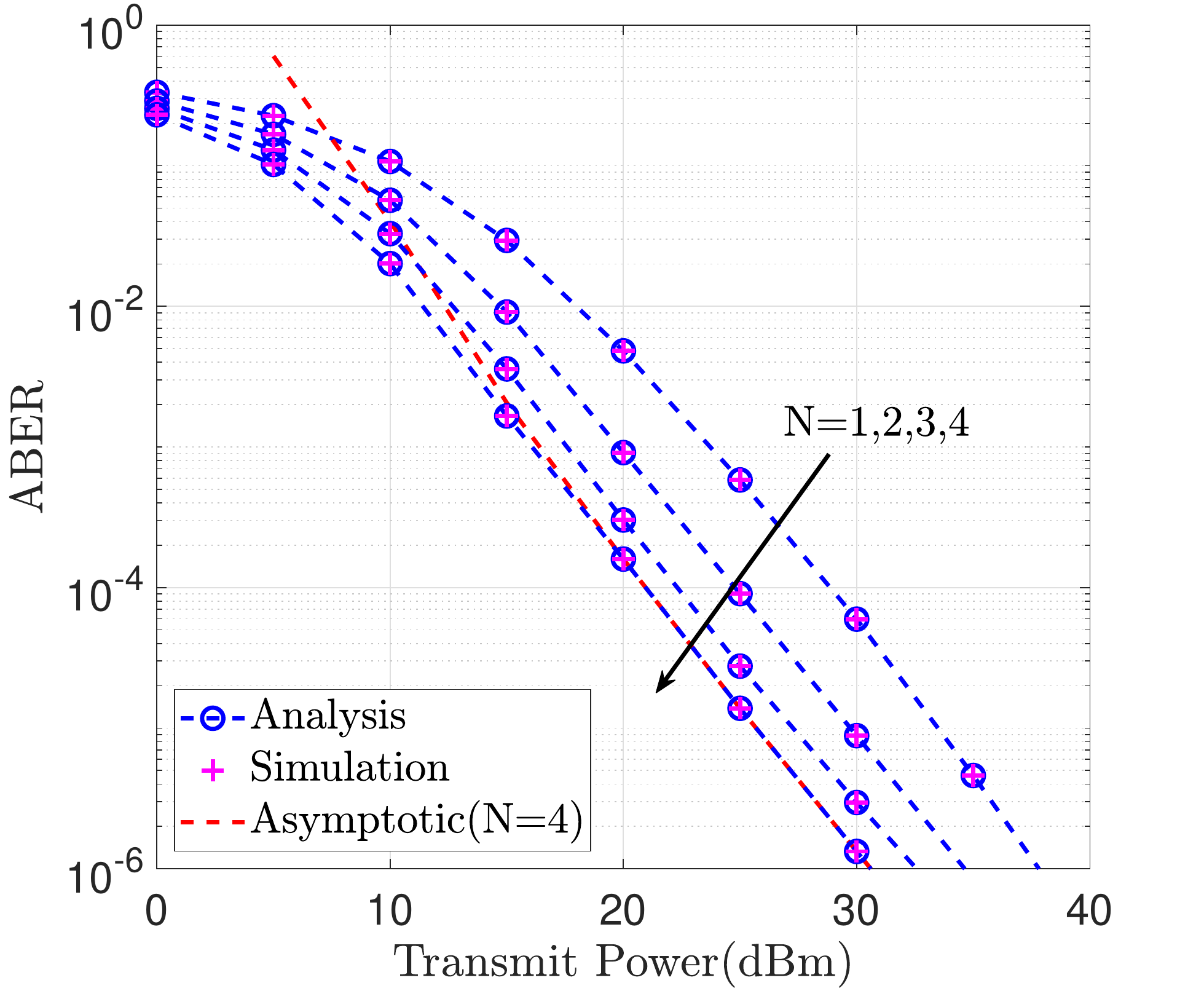}}
	\subfigure[Downlink transmission.]{\includegraphics[scale= 0.43] {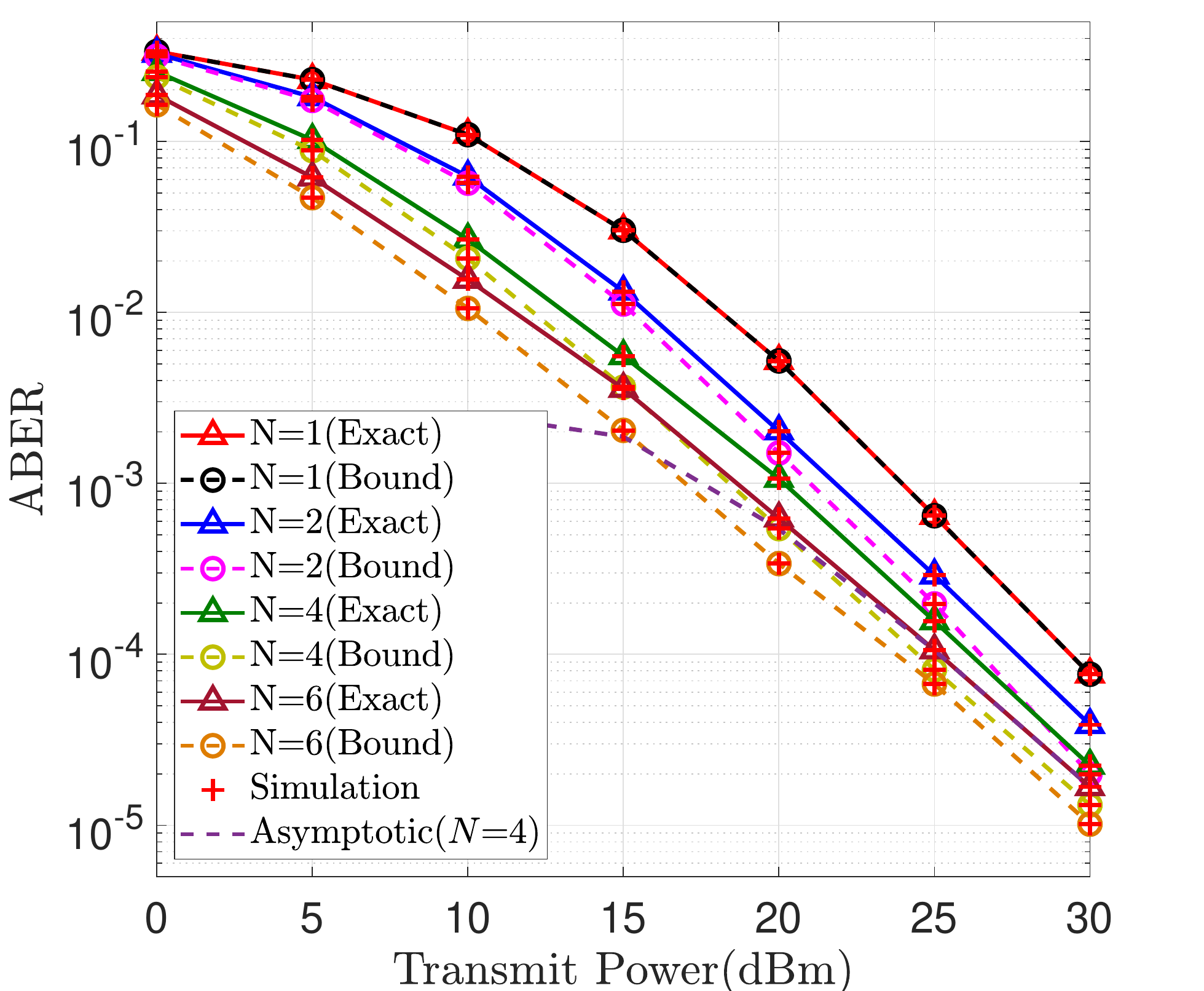}}
	\caption{Average BER of the mixed multihop relaying system with $\sigma_{dB}=5$, $\phi_i=37$, $d_{\rm MH}=50$ \mbox{m}, and $d_A=20$\mbox{m}.}
	\label{fig:ber}
\end{figure}

In Fig.~\ref{fig:ber_dual_hop}, we demonstrate the average BER performance of the mixed relaying system. We compare the average BER performance of the $N=2$ system (see Fig.~\ref{fig:ber_dual_hop}) with $N>2$ uplink  (see Fig.~\ref{fig:ber}(a)) and downlink  transmission (see Fig.~\ref{fig:ber}(b)). Comparing Fig.~\ref{fig:ber_dual_hop} and \ref{fig:ber} with the outage probability in Fig.~\ref{fig:outage_dual_hop} and \ref{fig:outage}, we can draw similar conclusions since the average BER follows a similar trend as that of the outage probability with for the same network configuration. We investigate the impact of short-term fading $m_g$ of the access link on the average BER. Fig.~\ref{fig:ber_dual_hop} shows that the average BER significantly degraded when the factor $m_g$ is reduced from $5$ to $1$. This happens due to a decrease in the multipath clustering, thereby worsening the channel condition. We can compare Fig.~\ref{fig:ber}(a) and Fig.~\ref{fig:ber}(b) with Fig.~\ref{fig:ber_dual_hop} to infer that multihop relaying with additional $3$ relays (i.e., $N=4$) in the backhaul extends the communication link to $70$\mbox{m} achieving the similar BER performance with $N=1$ system at a total link distance of $40$\mbox{m}. Further, the diversity order using average BER is  $M_{{\bar{P}_e}} = \frac{\alpha\mu}{2}=0.75$ when $\sigma_{\rm dB}=5$ (i.e., $m_\Omega = 2.5$) with $\alpha=1.5$ and $\mu=1$ since the slope of the plots in Fig.~\ref{fig:ber_dual_hop} does not change with variation in other channel parameters. However, with $\alpha=2$ and $\mu=2$, the BER-diversity order becomes $M_{{\bar{P}_e}}= 1.15$ when $m_g= 5$ and becomes $M_{{\bar{P}_e}}= 3.25$ when $m_g= 1$. The slope of the plots in Fig.~\ref{fig:ber_dual_hop} and \ref{fig:ber} clearly demonstrate the impact of pointing errors, fading parameters, and number of relays on the average BER performance.

In all the above plots, we validated the numerical evaluation of analytical and asymptotic expressions with the simulation results. The figures show that the numerical computation of the derived analytical expressions has an excellent agreement with the simulations for the CA-multihop and uplink systems. However, there exists a gap between the derived upper bound and simulation results when the number of relays becomes large for the FG-multihop and downlink systems. Note that the derived expressions are exact for $N=2$  uplink and downlink transmission systems.
\section{Conclusion and Future Work}\label{sec:conc}
In this paper, we evaluated the performance of an $N$+$1$-hop THz system consisting of an $N$-hop backhaul link and a single-hop shadowed radio access link over the THz band. We modeled the backhaul with generalized i.ni.d $\alpha$-$\mu$ fading and the access link with a composite generalized-$K$ shadowed fading channel, combined with pointing errors and path loss due to the atmospheric turbulence. We developed statistical results of the mixed $N+1$-hop link by deriving the exact and upper bounds of the PDF and CDF of the SNR for the CA-multihop and FG-multihop systems integrated with shadowed access link using DF and FG relaying protocols for uplink and downlink transmissions, respectively. We analyzed the impact of multiple relays in the backhaul link for extending the communication range using outage probability and average BER performance the mixed transmission system. The diversity order shows that a sufficient number of relays can mitigate the effect of fading and pointing errors in the multihop-assisted backhaul, limiting the high SNR performance of the mixed system on the channel fading of the single-hop access link when a sufficient beam-width is used in the access link for signal transmission. Extensive simulation and numerical analysis are presented to show that $4$ relays are required to extend the backhaul range from $20$ \mbox{m} to $50$ \mbox{m} using CA-assisted multihop relaying while few more relays are required for the simpler FG-assisted multihop due to the folding of background noise from previous hops. The cumulative gain of the AF-assisted multihop system reduces with an increase in the number of hops advocating the use of a limited number of relays. Further, the simpler FG-multihop transmission performs close to the CA-multihop when the non-linear characteristic of the THz channel  increases. We envision that the proposed analysis can provide several design and deployment criteria for THz based ultra-high-speed wireless networks for high-end applications. The future scope of the work can include the impact of hardware impairment for multihop THz transmissions.

\section* {Appendix A: Proof of Theorem 1}
Substituting \eqref{eq:pdf_hpf_Meijer} in \eqref{eq:moment_mh_eqn2} and solving the inner integral $ \int_{0}^{\infty} e^{-s/\gamma} \gamma^{\frac{\phi_i+\alpha_iS_1}{2}-1} d\gamma = s^{\frac{\phi_i+\alpha_iS_1}{2}} \Gamma\big(-\frac{\phi_i}{2}-\frac{\alpha_iS_1}{2}\big) $ with the definition of Fox's H-function \cite{Mathai_2010}, we get
\begin{flalign} \label{eq:mgf_single_link}
	&\mathcal{M}_{\frac{1}{\gamma_i}}(s)\hspace{-1mm} = \hspace{-1mm}\frac{A_is^{\frac{\phi_i}{2}}}{2{\gamma_N^0}^{\frac{\phi_i}{2}}}H_{1,3}^{3,0}\Bigg[\frac{C_is^{\frac{\alpha_i}{2}}}{{\gamma_N^0}^{\frac{\alpha_i}{2}}} \Bigg| \begin{matrix} (1,1) \\ (B_i,1), (0,1), \big(\frac{-\phi_i}{2}, \frac{\alpha_i}{2}\big) \end{matrix} \Bigg]
\end{flalign}
Using \eqref{eq:mgf_single_link} and \eqref{eq:moment_mh_eqn}  in \eqref{eq:cdf_CSI_exact_main_eqn} and applying the inverse Laplace transform with expansion of the Fox's H-function of \eqref{eq:mgf_single_link} in terms of Mellin-Barnes integral, we get

\begin{flalign}\label{eq:ca_zaf}
	F_{\gamma_N}^{\rm CA} (\gamma) =& 1- \prod_{i=1}^{N} \frac{A_i s^{\frac{\phi_i}{2}}}{2s{\gamma_N^0}^{\frac{\phi_i}{2}}}  \frac{1}{2\pi i} \int_{\mathcal{L}}^{} \Bigg[\Big(\frac{1}{2\pi i}\Big)^N \int_{{\mathcal{L}_i}}^{}\frac{\prod_{i=1}^{N} \Gamma(B_i-S_i)\Gamma(0-S_i)\Gamma\big(-\frac{\phi_i}{2}-\frac{\alpha_iS_i}{2}\big)}{\prod_{i=1}^{}\Gamma(1-S_i)} \nonumber \\ \times & \bigg(\prod_{i=1}^{N}\frac{C_i s^{\frac{\alpha_i}{2}}}{{\gamma_N^0}^{\frac{\alpha_i}{2}}}\bigg)^{S_i} dS_i\Bigg] e^{s\gamma} ds
\end{flalign}

where $\mathcal{L}$ is contour integral for the Laplace inverse transform and $\mathcal{L}_i$ is the the $i$-th Mellin-Barnes integral. Changing the order of the integrals and solving the the inner integral \\ $\frac{1}{2\pi i} \int_{\mathcal{L}}^{} s^{\big(\sum_{i=1}^{N} \frac{\phi_i+\alpha_iS_i}{2}-1\big)} e^{s\gamma} ds$ using \cite[8.315]{Gradshteyn}, \eqref{eq:ca_zaf} can be represented as
\begin{eqnarray} \label{eq:cdf_CSI_mh_int}
	&F_{\gamma_N}^{\rm CA} (\gamma)=1- \prod_{i=1}^{N} \frac{A_i}{2  {\gamma_N^0}^{\frac{\phi_i}{2}}} {\gamma}^{\big(\sum_{i=1}^{N}\frac{\phi_i}{2}\big)}\Bigg[\Big(\frac{1}{2\pi i}\Big)^N \int_{{\mathcal{L}_i}}^{} \frac{\prod_{i=1}^{N}\Gamma(B_i-S_i)\Gamma(0-S_i)\Gamma\big(-\frac{\phi_i}{2}-\frac{\alpha_iS_i}{2}\big)}{\prod_{i=1}^{N} \Gamma(1-S_i)} \nonumber\\  & \times \bigg(\prod_{i=1}^{N}\frac{C_i\gamma^{\frac{\alpha_i}{2}}}{{\gamma_N^0}^{\frac{\alpha_i}{2}} }\bigg)^{S_i} \frac{1}{\Gamma\big(-\sum_{i=1}^{N}\frac{\phi_i+\alpha_iS_i}{2}+1 \big)} dS_i\Bigg]   
\end{eqnarray}
We apply the definition of multivariate Fox's H-function \cite{Mathai_2010} to get the CDF in \eqref{eq:cdf_CSI_mh}. To get the PDF, we differentiate \eqref{eq:cdf_CSI_mh_int} with respect to $\gamma$ to get
\begin{eqnarray} \label{eq:pdf_CSI_mh_int}
	&f_{\gamma_N}^{\rm CA}(\gamma) = \Bigg[1-\prod_{i=1}^{N} \frac{A_i {\gamma^0}^{-\frac{\phi_i}{2}}}{2}    \bigg[\Big(\frac{1}{2\pi i}\Big)^N \int_{\mathcal{L}_i}^{} \frac{\prod_{i=1}^{N} \Gamma(B_i-S_i) \Gamma(0-S_i)\Gamma\big(-\frac{\phi_i}{2}-\frac{\alpha_iS_i}{2}\big)}{\prod_{i=1}^{N} \Gamma(1-S_i)} \nonumber \\  & \times \bigg(\hspace{-1mm}\prod_{i=1}^{N}\frac{ C_i}{{\gamma^0}^{\frac{\alpha_i}{2}}}\hspace{-1mm}\bigg)^{\hspace{-1mm}S_i} \hspace{-2mm} \frac{1}{\Gamma\big(-\sum_{i=1}^{N}\frac{\phi_i+\alpha_iS_i}{2}+1 \big)} dS_i\bigg]D_1 \Bigg]   
\end{eqnarray}
where $D_1=\frac{d}{d\gamma} {\gamma}^{\big(\sum_{i=1}^{N}\frac{\phi_i+\alpha_iS_i}{2}\big)}= \sum_{i=1}^{N}\frac{\phi_i+\alpha_iS_i}{2} {\gamma}^{\big(\sum_{i=1}^{N}\frac{\phi_i+\alpha_iS_i}{2}-1\big)}$. Hence,  applying the definition of multivariate Fox's H-function in \eqref{eq:pdf_CSI_mh_int}, we obtain the PDF in \eqref{eq:pdf_CSI_mh} to conclude the proof.

\section* {Appendix B: Proof of Theorem 2}
We use \eqref{eq:pzaf2} in \eqref{eq:gen_prod_pdf_1_mellin} to represent the PDF of AF-assisted multihop link as
\begin{eqnarray} \label{eq:pdf_multihop_int}
	f_{\gamma_N}^{\rm FG}(\gamma) = \prod_{i=1}^{N} \frac {\zeta_i A_i{C_i^{-\frac{\phi_i}{\alpha_i}}} } {N\alpha_i\gamma}  \frac{1}{2\pi \J} \int_{\mathcal{L}} \prod_{i=1}^{N} \bigg(\hspace{-1mm}{C_i^{\frac{2 l_i}{N\alpha_i}} {\gamma_N^0}^{-\frac{ l_i}{N}}}\gamma \hspace{-1mm}\bigg)^{\hspace{-1mm}-r} \frac{ \prod_{i=1}^{N} \Gamma\big( {\frac{2r l_i}{N\alpha_i}}+\frac{B_i\alpha_i+\phi_i}{\alpha_i}\big)\Gamma\big({\frac{2rl_i}{N\alpha_i}}+\frac{\phi_i}{\alpha_i}\big) }{ \prod_{i=1}^{N} \Gamma\big( {\frac{2r l_i}{N\alpha_i}} +\frac{\phi_i+\alpha_i}{\alpha_i} \big)} \diff r 
\end{eqnarray}

Finally, we apply the  definition of Fox's H-function \cite{Kilbas_2004} to get the PDF of $\gamma_N$ in \eqref{eq:pdf_multihop}. Further, we use \eqref{eq:pdf_multihop_int} in $F_{\gamma_N}(\gamma) = \int_{0}^{\gamma} f_{\gamma_N}(\gamma) d\gamma $, and solve the resultant  inner integral $\int_{0}^{\gamma} \gamma^{-r-1} d\gamma$ to $\frac{\gamma^{-r}}{-r} = \frac{\gamma^{-r} \Gamma(-r)}{\Gamma(1-r)} $ in order to apply the definition of Fox's H-function \cite{Mathai_2010} to represent the CDF in \eqref{eq:cdf_multihop}, which concludes the proof of Theorem 2.

\section* {Appendix C: Proof of Proposition 1}
Transforming random variable $h_p$ with PDF of \eqref{eq:pdf_hp} for the access link as $z=h_p^2$, we get 
\begin{eqnarray} \label{eq:pdf_hp_square}
	f_{z}(x)=\frac{1}{2} {\phi_A ^{2}S_A^{-\phi_A ^{2}}}{}x^{{\frac{\phi_A ^{2}}{2}-1}}, \quad 0 \leq x \leq {S_A^{2}},
\end{eqnarray}	  
Using the limits of PDF in \eqref{eq:pdf_hk} and \eqref{eq:pdf_hp_square}, the PDF of $|h_{A}|^{2}= h_{k}h_{p}^{2}$ is given by \cite{papoulis_2002}:
\begin{align}  \label{eq:meij_app_zaf}
	&f_{h_A^2}(x)\hspace{-1mm} =\hspace{-1mm} \frac{b^{m_{A}} x^{\frac{m_{A}}{2}-1} \phi_A S_A^2 }{\Gamma(m_{\Omega}) \Gamma(m_{g})} \hspace{-1mm} \int_{0}^{S_A^2} \hspace{-2mm} y^{\big(\frac{\phi_A -m_{A}}{2}-1\big)} K_{m_{M}} \Big(\hspace{-1mm}2b\sqrt{\frac{x}{y}}\Big) {d}y
\end{align}
{Using the Bessel function in terms of Meijer's G-function \cite[eq (14)]{Adamchik}, we represent \eqref{eq:meij_app_zaf} as
\begin{eqnarray} \label{eq:pdf_kp_combined_ZAF}
	f_{|{h_{kp}}|^2}(x) = \frac{b^{m_{A}} x^{\frac{m_{A}}{2}-1} \phi_A S_A^2 }{\Gamma(m_{\Omega}) \Gamma(m_{g})}  \int_{0}^{A_0^2} y^{\big(\frac{\phi_A -m_{A}}{2}-1\big)} G_{0,2}^{2,0}\left( {b^2 x y^{-1}} \Bigg| \begin{matrix} - \\ m_{M}, m_{M}\end{matrix} \right) {d}y
\end{eqnarray}
Further, we apply the integral representation of Meijer's G-function \cite{Mathai_2010} to express \eqref{eq:pdf_kp_combined_ZAF} as
\begin{eqnarray} \label{eq:pdf_kp_combined_int}
	f_{h_A^2}(x) = \frac{b^{m_{A}} x^{\frac{m_{A}}{2}-1} \phi_A (S_A^2)^{{\frac{\phi_A-m_{A}}{2}}+1} }{\Gamma(m_{\Omega}) \Gamma(m_{g})} \frac{1}{2\pi i} \int_{\mathcal{L}}^{}  \Gamma\big(m_{M}-s\big) \Gamma\big(m_{M}-s\big) (b^2 x)^s  I_{in}
\end{eqnarray}
where  $I_{in}=\int_{0}^{S_A^2}y^{{\frac{\phi_A-m_{A}}{2}}-1}y^{-s}ds=\frac{(S_A^2)^{{\frac{\phi_A-m_{A}-2s}{2}}}}{{{\frac{\phi_A-m_{A}-2s}{2}}}}=\frac{(S_A^2)^{{\frac{\phi_A-m_{A}}{2}}} (S_A^2)^{-s} \Gamma( {{\frac{\phi_A-m_{A}}{2}}}-s )} {\Gamma(1+ {{\frac{\phi_A-m_{A}}{2}}}-s ) }$. Finally, we apply the definition of Meijer's G-function in \eqref{eq:pdf_kp_combined_int} with the transformation of the random variable  $f_{\gamma_A}(\gamma) = \frac{1}{{\gamma_A^{0}}} f_{h_{A}}\big({\frac{\gamma}{\gamma{_A^0}}}\big) $ to get the PDF in \eqref{eq:pdf_kp_combined}. Substituting  \eqref{eq:pdf_kp_combined} in $\int_{0}^{\gamma}f_{{\gamma_{A}}}(\gamma) d\gamma $, and solving the definite integral involving a single Meijer's G-function, we get  CDF of SNR of  the shadowed access link.

\section* {Appendix D: Proof of Theorem 3}
Using \eqref{eq:pdf_hpf_Meijer} and \eqref{eq:pdf_kp_combined} in \eqref{eq:pdf_AF_eqn}, applying the definition of Meijer's G and Fox's H-functions, and changing the order of integration. we get
\begin{eqnarray} \label{eq:pdf_CG_int}
	&f_\gamma^{\rm DL}(z) = \Big[\prod_{i=1}^{N}\frac{\zeta_i A_i C_i^{-\frac{\phi_i}{\alpha_i}} b^{m_A}  \phi_A(S_A^2)^{\frac{\phi_A-m_A}{2}+1}z^{-1}}{2N\alpha_i{\gamma_A^0}^{\frac{m_{A}}{2}}\Gamma(m_{\Omega}) \Gamma(m_g)}\frac{1}{2\pi i}\int_{\mathcal{L}_1}^{} \frac{\prod_{i=1}^{N}\Gamma\big(\frac{B_i\alpha_i+\phi_i}{\alpha_i}+ \frac{2l_i}{N\alpha_i}S_1\big)\prod_{i=1}^{N}\Gamma\big(\frac{\phi_i}{\alpha_i}+\frac{2l_i}{N\alpha_i}S_1\big)}{\prod_{i=1}^{N}\Gamma\big(\frac{\phi_i+\alpha_i}{\alpha_i}+\frac{2l_i}{N\alpha_i}S_1\big)}\nonumber\\& \times \frac{1}{2\pi i}\int_{\mathcal{L}_2}^{}\frac{\Gamma(m_M-S_2)\Gamma(m_M-S_2)\Gamma\big(\frac{\phi_A-m_A}{2}-S_2\big)}{\Gamma\big(1+\frac{\phi_A-m_A}{2}-S_2\big)}\Bigg(\prod_{i=1}^{N}\frac{{{\gamma_i^0}^{\frac{l_i}{N}}}}{ C_i^{\frac{2 l_i}{k\alpha_i}}z}\Bigg)^{S_1} \bigg(\frac{b^2}{S_A^2 {\gamma_A^0}^{\frac{1}{2}}} \bigg)^{S_2} dS_1 dS_2\Big] I_{9}
\end{eqnarray}
where $I_9 = \int_{0}^{\infty} \big(\frac{x+\psi}{x}\big)^{-S_1} x^{\big(\frac{m_A-2+2S_2}{2}\big)} dx$ can be represented as $ \frac{\psi^{\frac{m_A}{2}} \psi^{S_2} \Gamma\big(-\frac{m_A}{2}- {S_2}\big) \Gamma\big(\frac{m_A}{2}+S_1+S_2\big) }{\Gamma(S_1)}$  \cite[3.241/4]{Gradshteyn}. Substituting $I_9$ in \eqref{eq:pdf_CG_int} and applying the definition of bivariate Fox's H-function \cite[1.1]{Mittal_1972}, we get the PDF of Theorem \ref{th:pdf_cdf_mixed_af} in \eqref{eq:pdf_CG}.

We use the PDF \eqref{eq:pdf_CG} in $\int_{0}^{\gamma} f_\gamma(z) d\gamma$ and solve the resultant integral applying the Mellin-Barnes integral representation of the Fox's H-function \cite[1.1]{Mittal_1972} to get the CDF in \eqref{eq:cdf_CG}, which concludes the proof.

\bibliographystyle{IEEEtran}
\bibliography{Thz_references_others}
\end{document}